\documentclass{article}

\usepackage{amsfonts}
\usepackage{amsmath}
\usepackage{amssymb}
\usepackage{amsthm}
\usepackage{breakcites}
\usepackage{caption}
\usepackage{enumitem}
\usepackage[a4paper, left=1in, right=1in, bottom=1.2in]{geometry}
\usepackage{graphicx}
\usepackage{hyperref}
\usepackage[utf8]{inputenc}
\usepackage{makecell}
\usepackage{multirow}
\usepackage{natbib}
\usepackage{siunitx}
\usepackage{subcaption}
\usepackage{tikz}

\graphicspath{{images/}}

\DeclareMathOperator*{\argmin}{arg\,min}
\newcommand{\Expval}[1]{\mathrm{E}\! \left(#1\right)}
\newcommand{\Var}[1]{\mathrm{Var}\! \left(#1\right)}
\newcommand{\us}{U_\mathrm{s}}
\newcommand{\up}{U_\mathrm{p}}
\newcommand{\xstar}{\mathbf{x}_*}
\newcommand{\repourl}{\href{https://github.com/llnl/BALSCD}{github.com/llnl/BALSCD}}

\renewcommand{\eqref}[1]{Eq.~\hyperref[#1]{(\ref*{#1})}}

\hypersetup{
  colorlinks,
  linkcolor={red!70!black},
  citecolor={blue!70!black},
  urlcolor={blue!80!black}
}

\begin{document}

\title{A Tutorial on Bayesian Analysis of \\Linear Shock Compression Data}

\author{
  Jason Bernstein$^{1}$\thanks{Corresponding author: bernstein8@llnl.gov}, 
  Philip C. Myint$^{1}$, 
  Beth A. Lindquist$^{2}$, 
  Justin Lee Brown$^{3}$ \\[1em]
  $^{1}$Lawrence Livermore National Laboratory, Livermore, CA 94550 \\
  $^{2}$Los Alamos National Laboratory, Los Alamos, NM 87545 \\
  $^{3}$Sandia National Laboratories, Albuquerque, NM 87185
}

\date{}

\maketitle

\begin{abstract}
Gas gun and other shock compression experiments often produce shock wave velocity measurements that are linearly associated with particle velocity.
Traditionally, this empirical relationship is quantified with a single Hugoniot curve that is estimated using least squares regression.
However, for downstream modeling and simulation tasks, it is often more useful to have multiple Hugoniot curves in the pressure-volume plane that are consistent with the data.
We employ Bayesian uncertainty quantification methods as a framework for propagating measurement uncertainty through to model parameters and predictions.
Specifically, this tutorial shows how to sample multiple Hugoniot curves in the pressure-volume plane that are consistent with the shock wave-particle velocity measurements in a two-step Bayesian approach.
First, we obtain an analytical expression for the posterior distribution of the linear model parameters using Bayesian linear regression.
Second, we propagate samples from the posterior distribution through the Rankine-Hugoniot equations to yield Hugoniot curves in the pressure-volume plane.
The procedure is demonstrated with publicly available data on argon, copper, and nickel, and compared against bootstrapping and linear regression.
The Bayesian procedure is shown to be interpretable, computationally inexpensive, and less sensitive than an alternative bootstrapping approach to the removal of the point in the copper dataset that has the largest particle velocity.
As a tutorial on Bayesian methodology for the shock compression community, we provide several derivations and explanations that make this paper self-contained, and make all code and data available at \repourl.
\end{abstract}

\section{Introduction}
\label{sec:introduction}

Shock compression experiments produce pairs of shock wave-particle velocity measurements at different impact velocities.
These data are valuable because it constrains the equation of state (EOS) of materials at extreme pressures and other conditions.
For many materials, the relationship between the shock wave and particle velocities appears to be linear, as~\cite{marsh1980} demonstrates for many datasets.
An empirically motivated linear model is therefore typically fit to relate the shock wave velocity measurements to the particle velocity measurements.
The goal of this tutorial is to show how to quantify uncertainty in the linear model parameters in order to produce Hugoniot curves in the shock wave-particle velocity plane that are consistent with the data.
These Hugoniot curves are then propagated through the Rankine-Hugoniot equations in order to obtain Hugoniot curves in the pressure-volume plane, which can assist in fitting parameters in EOS models, such as the Mie-Gr\"{u}neisen, Debye, Cell, and Cowan models~\citep{more1988,wu2021,myint2023}.
Tables generated from these EOS models can be used in hydrocode simulations~\citep{robinson2013}, and Hugoniot curves are also important for impedance matching~\citep[Ch. 3]{forbes2012}.
The challenges are, therefore, to accurately quantify the uncertainty in the linear model relating shock wave to particle velocity and to efficiently sample Hugoniot curves in the pressure-volume plane that are consistent with this uncertainty.

There is a growing body of literature on applications of uncertainty quantification (UQ) to materials at extreme conditions (such as under shock and dynamic compression), with a focus on Bayesian methods.
For background, uncertainty in the Bayesian paradigm is characterized by a probability distribution called the posterior distribution, which is obtained by combining a prior distribution over model parameters with the experimental measurements using Bayes' rule.
For non-linear models, Bayesian inference is often performed with an algorithm called Markov Chain Monte Carlo (MCMC), which draws samples of the model parameters from the posterior distribution.
For example, MCMC has been used to calibrate EOS models to shockless dynamic compression data~\citep{brown2023}, high-explosive (HE) reactant and product data~\citep{lindquist2023,lindquist2024b,lindquist2025,wang2025}, and linear shock wave-particle velocity data~\citep{robinson2013}.
\cite{schmid2025} compare MCMC and optimization for calibrating the Davis reactants EOS model.
MCMC has also been applied to calibrate the Johnson-Cook~\citep{walters2018} and Preston-Tonks-Wallace (PTW)~\citep{rivera2022} material strength models using Gaussian process surrogate models, jointly calibrate PTW and HE model parameters~\citep{nelms2024}, fit the Swegle-Grady scaling law for shock compression~\citep{gurrutxaga2025}, and calibrate a plasticity model for body-centered-cubic single crystals~\citep{lee2026}.
More broadly,~\cite{gaffney2022} train a Gaussian process that respects physical constraints on plasma EOS data, and~\cite{mentzer2023} train a neural network emulator on EOS data.
Overall, these papers reflect a trend of calibrating increasingly complex models with limited data.

Another approach to UQ is called bootstrapping, which involves repeatedly resampling the original dataset to obtain new datasets and then fitting the model of interest to each new dataset.
An example application of bootstrapping to EOS model development can be found in~\cite{ali2020}.
Bayesian and bootstrap methods both provide distributions over the parameters of interest, though the Bayesian approach naturally incorporates prior information into the distribution as we discuss later.
The two methods have also been combined for calibrating an EOS model from limited simulated and experimental data~\citep{lindquist2024}.

To the best of our knowledge, the literature does not contain a tutorial on Bayesian analysis of linear shock compression data or a closed-form Bayesian analysis.
Our analysis relies on the fact that if a certain prior distribution on the linear model and measurement error parameters is assumed, then the marginal posterior distribution of the linear model parameters is a bivariate $t$-distribution and this distribution can be sampled directly, removing the need for MCMC.
This result is well-known in the statistics literature~-~details can be found in~\citet[Ch. 14]{gelman1995},~\citet[Ch. 11]{rencher2008}, or~\cite{banerjee2008}, for example, and we describe the multivariate $t$-distribution in~\ref{app:tdistribution}~-~but it does not appear to have been applied to quantifying uncertainty in shock wave EOS.
Alternatively, MCMC can be used to sample from the posterior distribution of the linear model parameters, but this approach is more computationally expensive and does not provide the mean and covariance matrix of the posterior distribution in closed form.
Bootstrapping is another option for performing uncertainty quantification of linear models, and we provide a comparison of the Bayesian and bootstrapping approaches for completeness.
One difference is that the Bayesian analysis produces a distribution that quantifies uncertainty in the linear model parameters for the available dataset, whereas bootstrapping produces a distribution that quantifies uncertainty in the parameter estimates over different datasets.
Linear regression is also an option for analyzing linear shock compression data;
~\cite{marsh1980} effectively takes this approach by reporting least squares estimates of the linear model parameters.
As we will discuss, many of the linear regression estimates coincide with Bayesian estimates if a certain prior distribution is assumed, though the Bayesian modeling framework is more general due, in part, to the user-specified prior distribution.

Overall, this tutorial covers the following topics arising in shock compression analysis:
\begin{enumerate}
\item We show how to perform a Bayesian analysis of linear shock wave-particle velocity data, including inference of the linear model parameters and validation of the model.
To make the tutorial self-contained, we provide a derivation of the previously mentioned posterior bivariate $t$-distribution of the coefficients of the linear relation between shock wave and particle velocity.
\item We show how to sample the posterior distribution of the linear model parameters and how to propagate those samples through the Rankine-Hugoniot equations to obtain Hugoniot curves in the pressure-volume plane.
\item We apply this methodology to published data on argon, copper, and nickel from~\cite{marsh1980}.
\item We provide a detailed example comparing this methodology to bootstrapping.
\end{enumerate}
Furthermore, the plots and tables in this paper can be reproduced with the code and data in the repository \repourl.
The repo name, BALSCD, is an acronym that stands for Bayesian Analysis of Linear Shock Compression Data.
This repository also includes a script that can be used to perform the Bayesian and bootstrap analyses on new datasets.
For example, the script can be used to analyze datasets on pyrolusite, serpentine, and toluene from~\cite{marsh1980} that are contained in the repository but are not analyzed in this paper.
Additional shock wave-particle velocity datasets can be found in~\cite{marsh1980} or~\cite{levashov2004}.

We emphasize that Bayesian analysis of linear models is well-established, so this paper does not develop new methodology or provide novel statistical content;
instead, it aims to be a tutorial on the application of Bayesian modeling and calibration to analyzing linear shock compression data.
To summarize the Bayesian analysis, the posterior distribution of the linear model parameters is first identified as a bivariate $t$-distribution centered at the least squares estimate.
Samples are then drawn from this $t$-distribution and used to compute the shock wave-particle velocity Hugoniot curve, which is, in turn, used to solve the Rankine-Hugoniot equations for pressure and volume.
The method is shown to produce Hugoniot curves in the pressure-volume plane that are consistent with experimental measurements and to be less sensitive than bootstrapping to the removal of the point in the copper dataset that has the largest particle velocity.
This tutorial is not a full uncertainty quantification analysis of linear models for shock compression data;
for example, we do not consider the issue of model form error.
Instead, its focus is on presenting Bayesian calibration of the linear model parameters in a pedagogical format.

The outline of this paper is as follows.
Section~\ref{sec:motivation} presents the motivating problem in more detail and introduces the data that are analyzed throughout this paper.
Section~\ref{sec:blr} shows how to fit these data in a Bayesian framework.
Section~\ref{sec:uncprop} shows how to use the results from the Bayesian analysis to sample Hugoniot curves in the pressure-volume plane.
Section~\ref{sec:comparison} compares the Bayesian approach with bootstrapping and linear regression.
Section~\ref{sec:informative_prior} extends the Bayesian approach presented in Sec.~\ref{sec:blr} by considering informative prior distributions, which are used to include information about model parameters that is independent of the calibration data.
Section~\ref{sec:discussion} discusses Bayesian methods more generally, offers guidance on choosing a method, and notes several topics not covered in this tutorial.
Section~\ref{sec:conclusion} concludes the paper and suggests directions for future work.
We emphasize that all of the figures and tables in this paper can be created by running the code in the accompanying repository.
Reading the code in parallel with this paper, specifically the script \verb+recreate_paper_results.py+, can reinforce the main ideas and notation.

\section{Motivating Problem}
\label{sec:motivation}

\begin{figure}[h]
\centering
\begin{tikzpicture}
\node[draw, fill=brown!30, minimum width=2cm, minimum height=2cm] (box1) at (0,0) {};
\node[draw, fill=cyan!30, minimum width=2cm, minimum height=2cm] (box2) at (2,0) {};
\draw[->, thick] ([xshift=0cm]box2.west) -- ([xshift=1cm]box1.east) node[midway, above] {$U_\mathrm{s}$};
\fill ([xshift=-1cm, yshift=3mm]box2.west) circle (2pt);
\draw[->, thick]
([xshift=-1cm, yshift=3mm]box2.west) --
([xshift=-0.4cm, yshift=3mm]box1.east)
node[midway, above] {$U_\mathrm{p}$};
\fill ([xshift=-1.5cm, yshift=-5mm]box2.west) circle (2pt);
\draw[->, thick]
([xshift=-1.5cm, yshift=-5mm]box2.west) --
([xshift=-0.9cm, yshift=-5mm]box1.east)
node[midway, above] {$U_\mathrm{p}$};
\end{tikzpicture}
\caption{Illustration of a shock compression experiment where the particle velocity is $\up$ and the shock wave velocity is $\us$. The centerline represents the shock front and the brown and blue regions are the material behind and in front of the shock front, respectively. The black circles represent the material behind the shock front and the arrows indicate the velocities.}
\label{fig:shock_compression}
\end{figure}
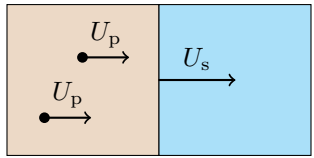

In a shock compression experiment, a shock wave moves through a material with velocity $\us$ and the material behind the shock front moves with velocity $\up$, as illustrated in Fig.~\ref{fig:shock_compression}.
Each experiment produces a single shock wave and particle velocity measurement, the latter being the velocity of the material behind the shock front.
In words, our goal is to perform uncertainty quantification on the parameters of a linear Hugoniot model relating shock wave velocity to particle velocity.
The linear model is usually written as
\begin{equation}
\label{hugoniot}
\us=C_0+S\up,
\end{equation}
where the intercept, $C_0$, is the bulk sound velocity and the slope, $S$, is the instantaneous rate of change of the shock wave velocity with the particle velocity.
Hence, $C_0$ has the same units as $\us$ and $\up$, and $S$ is unitless.
Note that not all shock wave-particle velocity datasets display a linear relationship;
for example,~\cite{ahrens1993} notes that the linear model is used when the material does not undergo a significant phase transition.
However, our analysis only considers datasets that are linear.
Additional details on shock compression physics and experiments can be found in~\cite{nellis1990},~\cite{ahrens1993},~\cite{forbes2012}, and~\cite{marsh1980}.

In order to perform a statistical analysis of the linear model in~\eqref{hugoniot}, we rewrite the model using standard linear model notation and include a measurement error term.
Assume there are $n$ shock wave and particle velocity measurements, denoted $Y_1,Y_2,\ldots,Y_n$ and $x_1,x_2,\ldots,x_n$, respectively, with $n\geq3$.
Each point comes from a separate experiment.
The published datasets do not provide separate uncertainties for the shock wave and particle velocities, so we consider a single measurement error term on the shock wave velocity that effectively captures all sources of experimental uncertainty.
A linear statistical model for the relationship between the shock wave and particle velocities is therefore
\begin{equation}
\label{linmod}
Y_i=C_0+S x_i+\epsilon_i
\end{equation}
for $i=1,\ldots,n$, where the $\epsilon_i$ are measurement errors that are modeled as independent and identically distributed normal random variables that have mean zero and variance $\sigma^2$;
the independence assumption is justified by the measurements coming from separate experiments.
This model can be rewritten in matrix notation as
\begin{equation}
\label{obsmodel}
Y=X\beta+\epsilon,
\end{equation}
where $\beta=(C_0,S)'$ denotes the Hugoniot parameters of interest, $Y=(Y_1,\ldots,Y_n)'$ denotes the shock wave velocity measurements, $X$ is an $n\times2$ design matrix whose first column is all ones and whose second column is $(x_1,\ldots,x_n)'$, and $\epsilon=(\epsilon_1,\ldots,\epsilon_n)'$.
The main statistical notation used in this paper is summarized in Table~\ref{tab:notation}.

\begin{table}[h]
    \caption{Statistical notation.}
    \label{tab:notation}
    \centering
    \begin{tabular}{|l|l|}
          \hline
          \textbf{Notation} & \textbf{Meaning} \\
          \hline
          $\beta=(C_0,S)'$ & Hugoniot parameters \\
          $\hat{\beta}=(\hat{C}_0,\hat{S})'$ & Estimate of $\beta$ \\ 
          $n$ & Number of measurements \\
          $Y=(Y_1,\ldots,Y_n)'$ & Shock wave measurements \\
          $x_i$ & $i$th particle velocity for $i=1,\ldots,n$ \\
          $X$ & Design matrix \\
          $\epsilon_i$ & $i$th measurement error \\
          $\sigma^2$ & Variance of measurement error \\
          $p(\beta,\sigma^2)$ & Prior distribution \\
          $p(\beta,\sigma^2|Y)$ & Joint posterior distribution of $\beta$ and $\sigma^2$ \\
          $p(\beta|Y)$ & Marginal posterior distribution of $\beta$ \\
          $s^2$ & Estimate of $\sigma^2$ \\
          $\Sigma=s^2(X'X)^{-1}$ & Scale matrix of posterior $t$-distribution \\
          $\nu=n-2$ & Degrees of freedom \\
          $t_2(\hat{\beta},\Sigma,\nu)$ & Posterior $t$-distribution of $\beta$ \\
          $\alpha\in(0,1)$ & Probability level \\
          $t_{\alpha,\nu}$ & $\alpha$th quantile of a univariate $t$-distribution \\
          $x_*$ & An arbitrary particle velocity \\
          $p(\tilde{y}|Y)$ & Posterior predictive distribution \\
          \hline
    \end{tabular}
\end{table}

The objective of our Bayesian analysis is to determine the probability distribution of the linear model parameters, $C_0$ and $S$, conditioned on the experimental measurements, or $p(\beta|Y)$.
As a by-product, we will also be able to infer the probability distribution of the measurement error variance, $\sigma^2$, conditioned on the experimental measurements, or $p(\sigma^2|Y)$.
These conditional probability distributions characterize the uncertainty in the model parameters given the experimental data, and can be used to answer questions such as ``Given our data, what is the probability that $S$ is contained in the interval (1.5, 1.6)?''.

\begin{figure}[h!]
  \centering
  \begin{subfigure}[b]{0.32\textwidth}
    \centering
    \includegraphics[width=\textwidth]{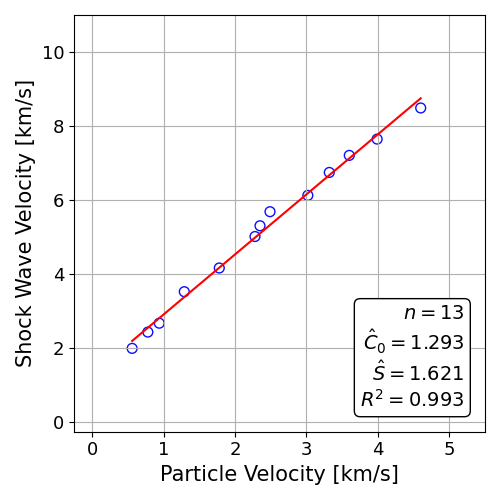}
    \caption{Argon}
  \end{subfigure}
  \begin{subfigure}[b]{0.32\textwidth}
      \centering
      \includegraphics[width=\textwidth]{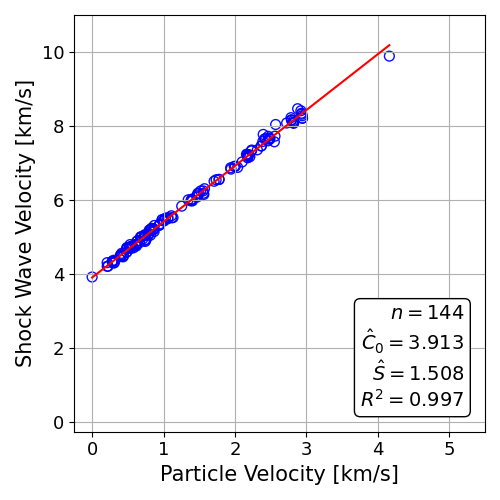}
      \caption{Copper}
      \label{fig:cu}
  \end{subfigure}
  \begin{subfigure}[b]{0.32\textwidth}
    \centering
    \includegraphics[width=\textwidth]{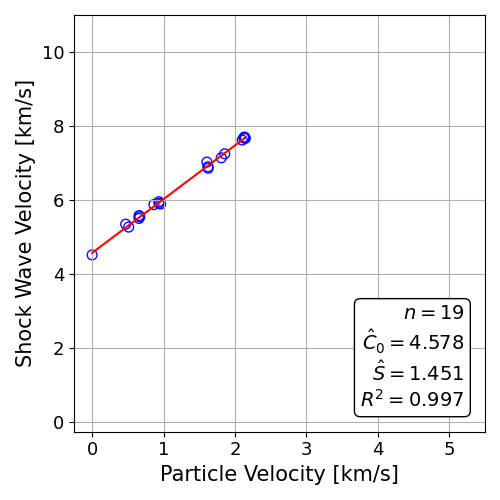}
    \caption{Nickel}
  \end{subfigure}
  \caption{Shock Hugoniot data for three materials from~\cite{marsh1980}. The lower right hand corner indicates the number of points, $n$, in each dataset, along with the least squares estimates of $C_0$ and $S$ and the $R^2$ statistic. Note that the axes are not drawn to scale.}
  \label{fig:data}
\end{figure}

We illustrate the Bayesian analysis using shock compression data on argon, copper, and nickel from~\cite{marsh1980}.
The data are shown in Fig.~\ref{fig:data} along with least squares fits.
These datasets were compiled from other sources that are referenced in~\cite{marsh1980}.
For the copper dataset $(n=144)$, 114 points are from impedance matching experiments, 26 are from shock and particle velocity experiments, 3 are from shock and free surface velocity experiments, and 1 is from a sound speed experiment.
The copper dataset originally had 115 points from impedance matching experiments, but we removed one duplicate point because it is unlikely that two experiments produced the exact same measurements.
The point with the largest particle velocity is from a shock and free surface velocity experiment.
All of the points in the argon dataset $(n=13)$ are from impedance matching experiments.
For the nickel dataset $(n=19)$, 18 of the points are from impedance matching experiments and 1 point is from a sound speed experiment.
The sound speed experiments are those with $\up=0$.
Note that the points generally fall on a straight line, but some points deviate from this line in a way that appears to be consistent with random noise or measurement error.

The linear model fits in Fig.~\ref{fig:data} are computed using the least squares estimate of $\beta$,
\begin{equation}
\label{ls_est_beta}
\hat{\beta}=(X'X)^{-1}X'Y,
\end{equation}
where $\hat{\beta}=(\hat{C}_0,\hat{S})'$.
This expression is derived in~\citet[Theorem 7.3a]{rencher2008} by minimizing the error sum of squares, or
\begin{equation}
\label{beta_hat}
\hat{\beta}=\argmin_\beta\,(Y-X\beta)'(Y-X\beta).
\end{equation}
The predicted shock wave velocity at a particle velocity $x_*$ is then given by $\xstar'\hat{\beta}$, where $\xstar=(1,x_*)'$.
The $R^2$ statistic shown in the plots is a measure of the strength of the linear correlation between the shock wave and particle velocities.
An $R^2$ equal to one would imply that shock wave velocity is a linear function of particle velocity, whereas a value of $R^2$ equal to zero would imply that there is no correlation between shock wave and particle velocity.
Hence, the $R^2$ values near one indicate that the two velocities are almost perfectly correlated.

A linear model can be fit to the shock Hugoniot data with least squares as shown in this section, but this model is limited since it assumes that the $\beta$ parameters are constants.
One way to incorporate prior information about $\beta$ into its estimate is to model $\beta$ with a probability distribution that reflects these beliefs;
for example, if it is believed that $C_0$ is between two and four, and close to three, then $C_0$ can be modeled as having a normal distribution that is centered at three and has a standard deviation of 0.5.
Inference of $\beta$ is then accomplished by updating this probability distribution with experimental data.
This procedure is called Bayesian inference because it relies on Bayes' rule, as described next.

\section{Bayesian Analysis}
\label{sec:blr}

The Bayesian linear regression model assumes that the observed shock wave velocities are generated from the linear model in~\eqref{obsmodel}, and is made Bayesian by assuming that $\beta$ and $\sigma^2$ are random variables.
There are two joint distributions over $\beta$ and $\sigma^2$ that are of interest.
First, there is the distribution $p(\beta,\sigma^2)$, which is called the prior distribution since it reflects beliefs about these parameters that are independent of the experimental data.
That is, the beliefs are formulated before the data are collected.
The second distribution of interest is $p(\beta,\sigma^2|Y)$, which is called the posterior distribution since it is the conditional distribution of $\beta$ and $\sigma^2$ given the shock wave velocity data.
This material and additional details on Bayesian analysis of linear models can be found in~\citet[Ch. 14]{gelman1995},~\citet[Ch. 11]{rencher2008}, or~\cite{banerjee2008}, for example.

Performing a Bayesian analysis requires specifying a prior distribution and computing the resulting posterior distribution.
By Bayes' rule, the posterior distribution is given by
\begin{equation}
\label{bayesrule}
p(\beta,\sigma^2|Y)=\frac{p(Y|\beta,\sigma^2)p(\beta,\sigma^2)}{p(Y)},
\end{equation}
where
\begin{equation}
p(Y)=\int p(Y|\beta,\sigma^2)p(\beta,\sigma^2)\,d\beta\,d\sigma^2
\end{equation}
is called the model evidence.
Note that the model evidence does not depend on $\beta$ and $\sigma^2$, the parameters of interest, since they are integrated out using the law of total probability.
Hence, we ignore the model evidence and focus instead on the two terms in the numerator of~\eqref{bayesrule}.
The first term follows from the Gaussian measurement model in~\eqref{obsmodel} as
\begin{equation}
\label{likelihood}
p(Y|\beta,\sigma^2)=\frac{1}{(2\pi\sigma^2)^{n/2}}\exp\left(-\frac{1}{2\sigma^2}(Y-X\beta)'(Y-X\beta)\right).
\end{equation}
This term is called the likelihood of the model when it is written as a function of $(\beta,\sigma^2)$, as in $L(\beta,\sigma^2)$, and captures the contribution of the data to the posterior distribution.
The second term is the prior distribution that is selected by the analyst.
We employ the prior distribution
\begin{equation}
\label{improper_prior}
p(\beta,\sigma^2)\propto \frac{1}{\sigma^2},
\end{equation}
which is called improper because it does not integrate to one and therefore is not an actual probability distribution.
However, this prior distribution can be used because it leads to a posterior distribution that integrates to one.
This prior distribution is commonly used for several reasons.
First, it does not depend on $\beta$, and therefore it does not bias the posterior distribution toward any particular values of $\beta$.
Second, the distribution does not encode strong beliefs about the measurement error variance;
see~\cite[Sec. 2.8]{gelman1995} for details.
Third, for linear models with likelihood given by~\eqref{likelihood}, this prior distribution leads to a posterior distribution that is available in closed form and can be sampled easily, as described next.
The prior distribution is said to be non-informative for the first two reasons.
We also provide an analysis with an informative, normal prior distribution on $\beta$ in Sec.~\ref{sec:informative_prior}.

The joint posterior distribution of the physics and measurement error parameters is derived in closed form in~\ref{app:joint_posterior} as
\begin{equation}
p(\beta,\sigma^2|Y)=p(\beta|\sigma^2,Y)p(\sigma^2|Y),
\end{equation}
where
\begin{equation}
\label{posterior_beta_given_sigmasq_Y}
p(\beta|\sigma^2,Y)=\frac{1}{2\pi\sigma^2|X'X|^{-1/2}}\exp\left(-\frac{1}{2\sigma^2}(\beta-\hat{\beta})'(X'X)(\beta-\hat{\beta})\right)
\end{equation}
and
\begin{equation}
\label{posterior_sigmasq}
p(\sigma^2|Y)=\frac{b^a}{\Gamma(a)}\frac{1}{(\sigma^2)^{a+1}}\exp\left(-\frac{b}{\sigma^2}\right)
\end{equation}
for $a=(n-2)/2$ and $b=(n-2)s^2/2$.
That is, the distribution of $\beta$ given $\sigma^2$ and $Y$ is a normal distribution with mean $\hat{\beta}$ and covariance matrix $\sigma^2(X'X)^{-1}$, denoted $\beta|\sigma^2,Y\sim N(\hat{\beta},\sigma^2(X'X)^{-1})$, and the marginal posterior distribution of $\sigma^2$ is an inverse gamma (IG) distribution with shape parameter $a$ and scale parameter $b$, denoted $\sigma^2|Y\sim IG(a,b)$.
This shows that the joint posterior distribution is a normal-inverse-gamma distribution.
Figure~\ref{fig:posterior_sigma_sq} in~\ref{app:marginal_posterior_sigsq} shows the posterior distribution of $\sigma^2$ for the datasets considered in this paper.

The interpretation of $p(\beta|\sigma^2,Y)$ is that if $\sigma^2$ is known, then the posterior distribution of $\beta$ is normal.
This hierarchical factorization of the joint posterior distribution suggests that one way to sample a $\beta$ and $\sigma^2$ from their joint posterior distribution is to first sample $\sigma^2\sim IG(a,b)$ following~\eqref{posterior_sigmasq} and then to sample ${\beta\sim N(\hat{\beta},\sigma^2(X'X)^{-1})}$ following~\eqref{posterior_beta_given_sigmasq_Y}.
Such samples of $\beta$ and $\sigma^2$ can be used for simulating synthetic data or for estimating integrals using Monte Carlo.
For example, a Monte Carlo estimate of the probability that $C_0$ is greater than 4, conditioned on the available data, is
\begin{equation}
p(C_0>4|Y)\approx M^{-1}\sum_{i=1}^M I\{C_0^{(i)}>4\},
\end{equation}
where $\beta^{(i)}=(C_0^{(i)},S^{(i)})'$ is the $i$th of $M$ samples of $\beta$ and $I\{\cdot\}$ is an indicator function.

Now, our goal is to obtain the marginal posterior distribution of $\beta$, the physics parameters of interest, without the nuisance parameter $\sigma^2$ related to the experimental measurements.
To this end, the marginal posterior distribution of $\beta$ is obtained by integrating $\sigma^2$ out of the full posterior distribution, $p(\beta,\sigma^2|Y)$, and is given in closed form as
\begin{equation}
\label{posterior_beta_pdf}
p(\beta|Y)=\frac{\Gamma(n/2)}{\Gamma(\nu/2)\nu\pi|\Sigma|^{1/2}}{\left[1+\frac{1}{\nu}(\beta-\hat{\beta})'\Sigma^{-1}(\beta-\hat{\beta})\right]}^{-n/2},
\end{equation}
which is the density of a bivariate $t$-distribution with $\nu=n-2$ degrees of freedom, mean $\hat{\beta}$, and scale matrix
\begin{equation}
\label{Sigma}
\Sigma=s^2(X'X)^{-1},
\end{equation}
where
\begin{equation}
s^2=\frac{1}{\nu} (Y-X\hat{\beta})'(Y-X\hat{\beta})
\end{equation}
is an estimate of the variance of the measurement errors and is called the sample variance;
see~\ref{app:tdistribution} for a description of the multivariate $t$-distribution and~\ref{app:marginal_posterior_beta} for a derivation of~\eqref{posterior_beta_pdf}.
The marginal posterior distribution of $\beta$ can also be expressed using the shorthand notation
\begin{equation}
\label{posterior_beta}
\beta|Y\sim t_2(\hat{\beta},\Sigma,\nu).
\end{equation}
The mean of the marginal posterior distribution of $\beta$, called the posterior mean, is also the least squares estimate of $\beta$, or
\begin{equation}
\label{expval_beta}
\Expval{\beta|Y}=\hat{\beta}.
\end{equation}
The covariance matrix of the marginal posterior distribution of $\beta$ is
\begin{equation}
\label{posterior_covar}
\Var{\beta|Y}=\frac{\nu}{\nu-2}\Sigma,
\end{equation}
which requires $n>4$ and is not the scale matrix, $\Sigma$, but closely approximates the scale matrix for large $n$.
The posterior mean and covariance matrix are also derived in~\ref{app:marginal_posterior_beta}.
Note also that $\hat{\beta}$ maximizes the posterior distribution in~\eqref{posterior_beta_pdf}, and so is called the maximum a posteriori (MAP) estimate of $\beta$ in a Bayesian context.
Overall, the posterior distribution of $\beta$ is in closed form, the posterior mean and variance of $\beta$ are available in closed form, and samples from the posterior distribution of $\beta$ can be obtained exactly.

\begin{figure}[h!]
  \centering
  \begin{subfigure}[b]{0.32\textwidth}
    \centering
    \includegraphics[width=\textwidth]{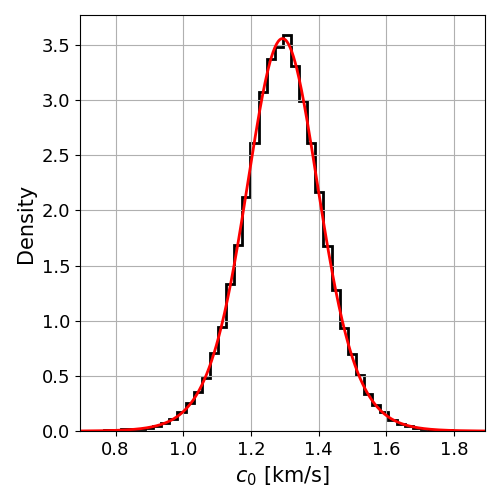}
    \caption{Argon}
  \end{subfigure}
  \begin{subfigure}[b]{0.32\textwidth}
      \centering
      \includegraphics[width=\textwidth]{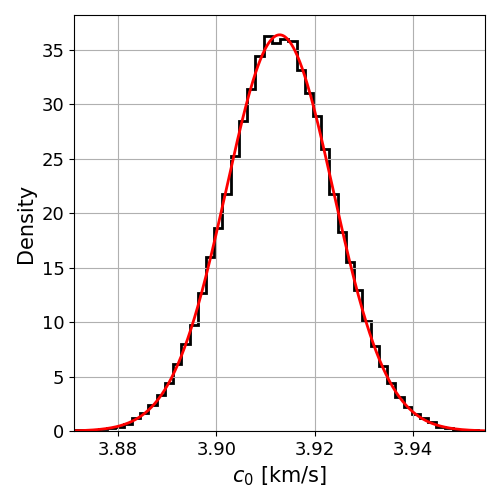}
      \caption{Copper}
  \end{subfigure}
  \begin{subfigure}[b]{0.32\textwidth}
    \centering
    \includegraphics[width=\textwidth]{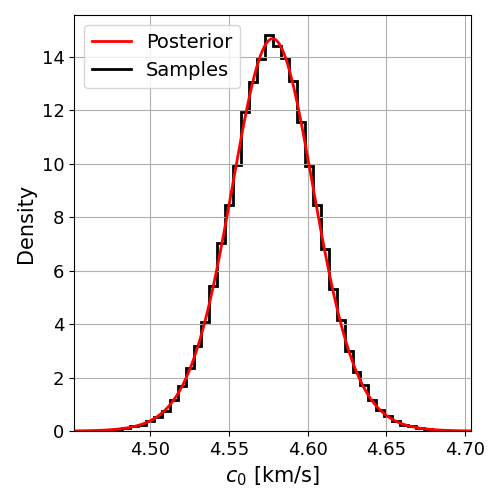}
    \caption{Nickel}
  \end{subfigure}
  \caption{The marginal posterior distribution of $C_0$ and a histogram of samples drawn from this distribution. The analytic distribution is useful because it fully characterizes the uncertainty in the Hugoniot model parameters, whereas the samples are useful for Monte Carlo and uncertainty propagation.}
  \label{fig:marginal_density}
\end{figure}

A key point of this tutorial is that the posterior distribution of $\beta$ is available in closed form and is given by~\eqref{posterior_beta}.
To demonstrate the utility of this result, Fig.~\ref{fig:marginal_density} shows the $t$-distribution of $C_0$ implied by~\eqref{posterior_beta} and a histogram of samples from this distribution obtained with the two-step algorithm given above.
As expected, there is close agreement between the analytic posterior distribution and histogram of the samples from the posterior distribution.
Both distributions are useful, though.
In particular, the analytic distribution provides closed-form expressions for the mean and covariance matrix of the $C_0$ and $S$ parameters, whereas the samples can be used for Monte Carlo calculations, as noted previously in this section.
Note also that in this and subsequent figures, the axis labels are $c_0$ or $s$ instead of $C_0$ or $S$.
The distinction is that $C_0$ and $S$ are random variables, whereas $c_0$ and $s$ refer to values these random variables can take on.

\begin{figure}[h!]
  \centering
  \begin{subfigure}[b]{0.32\textwidth}
    \centering
    \includegraphics[width=\textwidth]{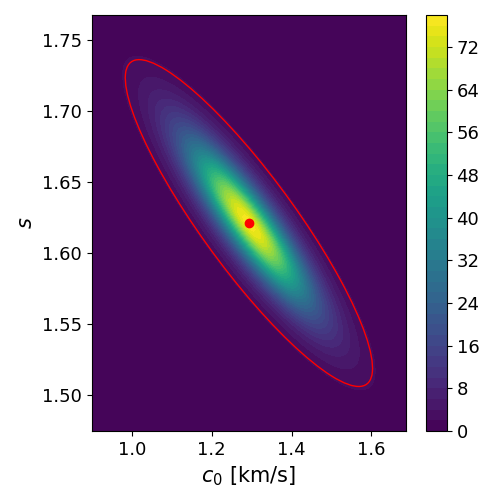}
    \caption{Argon}
  \end{subfigure}
  \begin{subfigure}[b]{0.32\textwidth}
      \centering
      \includegraphics[width=\textwidth]{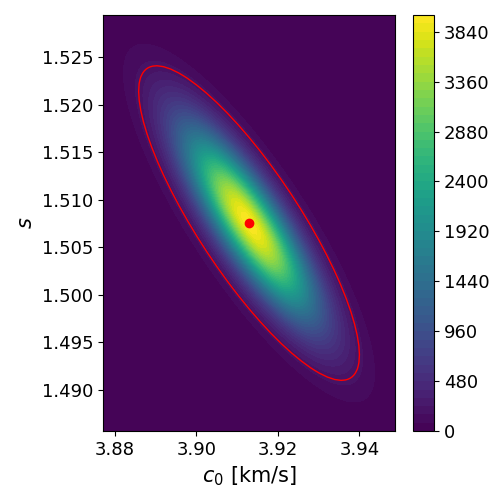}
      \caption{Copper}
  \end{subfigure}
  \begin{subfigure}[b]{0.32\textwidth}
    \centering
    \includegraphics[width=\textwidth]{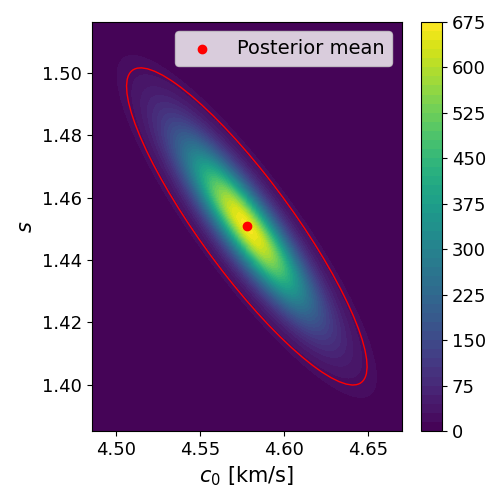}
    \caption{Nickel}
  \end{subfigure}
  \caption{Posterior distributions of $C_0$ and $S$. The red ellipses are 95\% credible regions, meaning they contain 95\% of the posterior probability.}
  \label{fig:posterior_beta}
\end{figure}

Figure~\ref{fig:posterior_beta} shows the posterior distribution of $\beta$ for the three datasets considered in this paper.
Observe that copper has the most concentrated posterior distribution, followed by nickel and then argon.
This occurs for two reasons.
First, the sample standard deviation is largest for argon $(s=0.182)$, then decreases for copper $(s=0.072)$ and nickel $(s=0.055)$. 
Second, the posterior variance tends to decrease as $n$ increases, and copper has the most observations, followed by nickel and then argon.
To see why the variance decreases with $n$, note that the $(X'X)^{-1}$ term in the posterior covariance matrix is~\citep[Ex. 7.3.2a]{rencher2008}
\begin{equation}
\label{inv_cov_matrix}
(X'X)^{-1}=\frac{1}{\sum_i(x_i-\bar{x})^2}\begin{pmatrix}
  n^{-1}\sum_i x_i^2 & -\bar{x} \\
  -\bar{x} & 1
  \end{pmatrix},
\end{equation}
where $x_i$ is the $i$th particle velocity and $\bar{x}$ is the mean of the $x_i$.
Now, the denominator of the fraction increases with $n$ while the non-constant terms in the matrix are means and therefore tend to stay approximately constant with $n$, so that overall the terms in this matrix trend toward zero.
Note also that the posterior correlation between $C_0$ and $S$ is negative since the off-diagonal terms in~\eqref{inv_cov_matrix} are $-\bar{x}$ and the mean particle velocity, $\bar{x}$, is positive.
Intuitively, the correlations are negative because if $C_0$ increases, then $S$ has to be decreased in order for the Hugoniot line to go through the shock wave-particle velocity measurements, and conversely if $C_0$ decreases, then $S$ has to be increased to compensate for a smaller intercept term.

The red ellipses in Fig.~\ref{fig:posterior_beta} are 95\% credible regions for $\beta$, or a set of values such that the probability that $\beta$ is contained in that region, conditioned on the experimental data, is 0.95.
The credible region for $\beta$ is elliptical by~\eqref{posterior_beta_pdf}.
A derivation of this credible region is given in~\ref{app:credible_region_beta}.

Similar to credible regions, an interval that contains an element of $\beta$ with a specified probability under the posterior probability distribution is called a credible interval.
For example, a 95\% credible interval for $C_0$ is
\begin{equation}
\label{credint_C0}
(\hat{C}_0 + t_{0.025,\nu}\sigma_1,\hat{C}_0 + t_{0.975,\nu}\sigma_1),
\end{equation}
where $\sigma_1^2$ is the $(1,1)$th element of $\Sigma$ and $t_{0.025,\nu}$ and $t_{0.975,\nu}$ are the 0.025th and 0.975th quantiles of a $t$-distribution with $\nu$ degrees of freedom.
Similarly, a 95\% credible interval for $S$ is
\begin{equation}
\label{credint_S}
(\hat{S} + t_{0.025,\nu}\sigma_2,\hat{S} + t_{0.975,\nu}\sigma_2),
\end{equation}
where $\sigma_2^2$ is the $(2,2)$th element of $\Sigma$.
In words, the probability that $C_0$ is contained in the interval given by~\eqref{credint_C0}, conditioned on the experimental data, is 0.95, and the same interpretation holds for $S$ and its credible interval.
These intervals are derived in~\ref{app:credible_intervals_C0_S} using the fact that the marginal posterior distributions of the elements of $\beta$ are univariate $t$-distributions.

For this two-dimensional posterior distribution, the univariate credible intervals are simpler to write down than the bivariate credible regions, but the credible regions are more informative since they capture the correlation between $C_0$ and $S$.

Table~\ref{tab:posterior_regression_coefs} summarizes the posterior distribution of the $C_0$ and $S$ parameters for the argon, copper, and nickel datasets.
The posterior mean is $\hat{\beta}$, the posterior standard deviations are given by the square roots of the diagonal elements of $\nu/(\nu-2)\Sigma$, and the credible intervals are given by~\eqref{credint_C0} and~\eqref{credint_S} for $C_0$ and $S$, respectively.
Note again the decrease in posterior variance as the sample size increases.

\begin{table}[ht]
\caption{Posterior means, standard deviations, and 95\% credible intervals for $C_0$ and $S$.}
\label{tab:posterior_regression_coefs}
\centering
\begin{tabular}{|l|l|c|c|c|c|}
\hline
\textbf{Material} & \textbf{Parameter} & \makecell[c]{\textbf{Posterior}\\\textbf{Mean}} & \makecell[c]{\textbf{Posterior Standard}\\\textbf{Deviation}} & \makecell[c]{\textbf{95\% Credible}\\\textbf{Interval}} & \textbf{Units} \\
\hline
\multirow{2}{*}{Argon}  & $C_0$ & $1.293$ & 0.121 & (1.052, 1.535) & \unit{km/s} \\
                        & $S$   & $1.621$ & 0.045 & (1.531, 1.711) & $\cdot$ \\
\hline
\multirow{2}{*}{Copper} & $C_0$ & $3.913$ & 0.011 & (3.891, 3.935) & \unit{km/s} \\
                        & $S$   & $1.508$ & 0.007 & (1.494, 1.521) & $\cdot$ \\
\hline
\multirow{2}{*}{Nickel} & $C_0$ & $4.578$ & 0.028 & (4.521, 4.634) & \unit{km/s}\\
                        & $S$   & $1.451$ & 0.020 & (1.411, 1.491) & $\cdot$ \\
\hline
\end{tabular}
\end{table}

This section gave an analytical expression for the posterior distribution of the $\beta$ parameters and showed how to summarize and draw samples from this distribution.
These samples can be propagated through the Rankine-Hugoniot equations to obtain Hugoniot curves in the pressure-volume plane.
The posterior distribution can also be propagated analytically to obtain the distribution of the shock wave velocity as a function of the particle velocity.
We continue in the next section with details on both uncertainty propagation tasks.

\section{Uncertainty Propagation}
\label{sec:uncprop}

This section covers two topics related to uncertainty propagation.
First, we show how to sample Hugoniot curves in the pressure-volume plane that are consistent with the experimental data by sampling $\beta$ from its marginal posterior distribution and propagating those samples through the non-linear Rankine-Hugoniot equations.
Second, we show how to characterize uncertainty in the mean shock wave velocity and future shock wave velocity measurements as a function of particle velocity.
The order of topics may seem backward because the previous section focused on shock wave-particle velocity data, but we chose this order because the first topic only requires information about the posterior distribution presented thus far, whereas the second topic requires a new concept called the posterior predictive distribution.

The Rankine-Hugoniot equations are a set of conservation equations that relate the shock wave and particle velocities to the change in pressure, volume, and internal energy of the material.
The conservation of mass equation is
\begin{equation}
\label{masscons}
\frac{V}{V_0}=\frac{\us-\up}{\us},
\end{equation}
the conservation of momentum equation is
\begin{equation}
\label{momentumcons}
P-P_0=\rho_0\us\up,
\end{equation}
and the conservation of energy equation is
\begin{equation}
\label{energycons}
E-E_0=\frac{1}{2}(P+P_0)(V_0-V),
\end{equation}
where $V_0$, $P_0$, $E_0$, and $\rho_0$ are the initial volume, pressure, internal energy, and density of the material ahead of the shock front, respectively, and $V$, $P$, and $E$ are the volume, pressure, and internal energy, respectively, of the material behind the shock front.
Hence, a Hugoniot curve in the pressure-volume plane can be obtained from these equations and a Hugoniot curve in the shock wave-particle velocity plane.

Hugoniot curves can be sampled in the pressure-volume plane with the following three-step procedure:
\begin{enumerate}
\item Following~\cite{hofert2013}, draw a sample from the marginal posterior distribution of $\beta$ defined in~\eqref{posterior_beta},
\begin{equation}
\label{tsample}
\beta=L\frac{Z}{\sqrt{W/\nu}}+\hat{\beta},
\end{equation}
where $Z=(Z_1,Z_2)'$ is a bivariate normal random vector with mean $(0,0)'$ and identity covariance matrix, $W$ is a chi-square random variable with $\nu=n-2$ degrees of freedom that is independent of $Z$, and $L$ is a lower-triangular matrix obtained from the Cholesky decomposition of $\Sigma$ and satisfies $\Sigma=LL'$.
In Python, this procedure for sampling $\beta$ is implemented in the \verb+multivariate_t+ class in the \verb+scipy.stats+ module.
\item Given this $\beta$, evaluate the corresponding Hugoniot curve in the shock wave-particle velocity plane over a grid of $\up$ values using~\eqref{hugoniot}.
\item Propagate the linear Hugoniot curve through the Rankine-Hugoniot equations,~\eqref{masscons} and~\eqref{momentumcons}, to obtain a Hugoniot curve in the pressure-volume plane.
\end{enumerate}
In words, these steps sample values of $C_0$ and $S$ from the posterior distribution and propagate those values through the Rankine-Hugoniot equations to obtain a Hugoniot in the pressure-volume plane.
The procedure for sampling $\beta$ from its marginal posterior distribution given in Sec.~\ref{sec:blr} could also be used for step 1, but the procedure given here is useful for deriving certain results later in this section.
Note also that given the sampled values of $P$ and $V$, the internal energy can be computed from~\eqref{energycons} if a value of $E_0$ is available.

\begin{figure}[h!]
  \centering
  \begin{subfigure}[b]{0.32\textwidth}
    \centering
    \includegraphics[width=\textwidth]{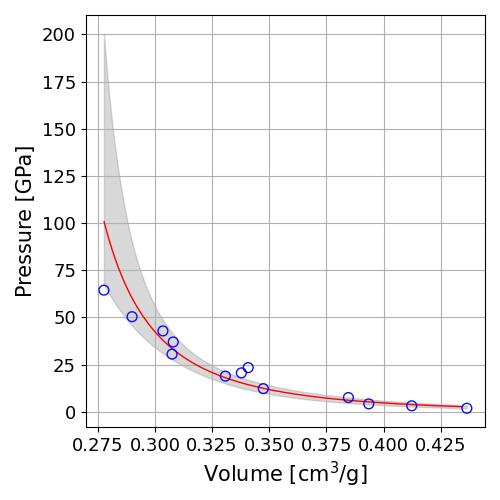}
    \caption{Argon}
  \end{subfigure}
  \begin{subfigure}[b]{0.32\textwidth}
      \centering
      \includegraphics[width=\textwidth]{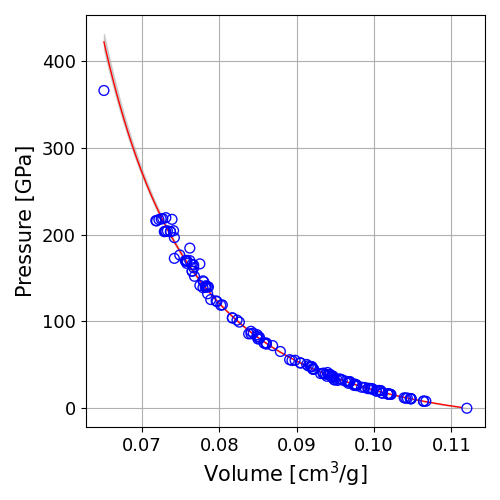}
      \caption{Copper}
  \end{subfigure}
  \begin{subfigure}[b]{0.32\textwidth}
    \centering
    \includegraphics[width=\textwidth]{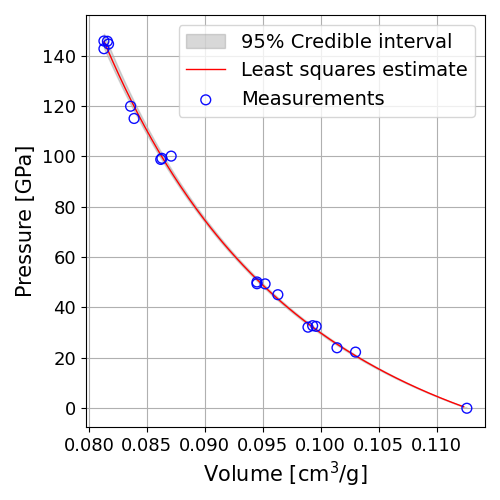}
    \caption{Nickel}
  \end{subfigure}
  \caption{Credible intervals for the Hugoniot curve in the pressure-volume plane, obtained from the posterior distribution of $C_0$ and $S$. The maximum widths of the intervals for the copper and nickel datasets are 18.3 GPa and 5.7 GPa, respectively, which makes these intervals difficult to see.}
  \label{fig:Hugoniot_samples}
\end{figure}

Figure~\ref{fig:Hugoniot_samples} shows 95\% credible intervals for pressure as a function of volume.
That is, the figure does not show individual Hugoniot curve samples, but rather a summary of their distribution.
To create this plot, we first sample a large number of $\beta$ vectors from their posterior distribution and then evaluate the corresponding shock wave-particle velocity curves over a grid of $\up$ values that includes the range of particle velocity measurements.
The Rankine-Hugoniot equations for volume and pressure given in~\eqref{masscons} and~\eqref{momentumcons}, respectively, are then evaluated taking the initial density, $\rho_0$, to be the average value of the initial densities in~\cite{marsh1980}, taking the initial volume to be one over this mean initial density, and taking the initial pressure, $P_0$, to be 1 bar, or $10^{-4}$ GPa, corresponding to ambient conditions.
At this point, we have a collection of curves in the pressure-volume plane, but the volumes are different for different samples of $\beta$, so we perform a linear interpolation step to adjust the Hugoniot curves to a common set of volumes.
Finally, we compute the 0.025th and 0.975th pressure quantiles across this grid of volumes to obtain the shaded regions in Fig.~\ref{fig:Hugoniot_samples}.
We see, for example, that the variability in pressure for argon can be on the order of 125 GPa for small volumes, but less than 5 GPa for large volumes;
the change in width of the credible intervals is due to the non-linearity of the Rankine-Hugoniot equations.
For the copper and nickel datasets, the variability in pressure is significantly less than for the argon dataset since their posterior distributions for $\beta$ are more concentrated, as seen in Fig.~\ref{fig:posterior_beta}.
Note also that the points with the highest particle and shock wave velocities have the smallest volumes and highest pressures in these plots, indicating the greatest degree of compression.

The posterior distribution of $\beta$ can also be propagated analytically through~\eqref{hugoniot} to quantify uncertainty in the mean shock wave velocity at a specified particle velocity.
In particular, the posterior distribution of the mean shock wave velocity at a particle velocity $x_*$ is derived by multiplying both sides of~\eqref{tsample} by $\xstar'=(1,x_*)$ to get
\begin{equation}
\xstar'\beta=\xstar'LZ/\sqrt{W/\nu}+\xstar'\hat{\beta}.
\end{equation}
By linearity, $\xstar'LZ$ is normally distributed with mean 0 and variance $\xstar'\Sigma\xstar$, and since every multivariate $t$-distribution can be expressed in the form of~\eqref{tsample} for different choices of the scale matrix, mean vector, and degrees of freedom, it follows that the posterior distribution of $\us=\xstar'\beta$ is a univariate $t$-distribution with mean $\xstar'\hat{\beta}$, scale parameter $\xstar'\Sigma\xstar$, and $\nu$ degrees of freedom, written
\begin{equation}
\label{post_dist_us}
\us|Y\sim t_1(\xstar'\hat{\beta},\xstar'\Sigma\xstar,\nu).
\end{equation}
A 95\% credible interval for the mean shock wave velocity is therefore
\begin{equation}
\label{credint_Us}
(\xstar'\hat{\beta}+t_{0.025,\nu}\sqrt{\xstar'\Sigma\xstar},\xstar'\hat{\beta}+t_{0.975,\nu}\sqrt{\xstar'\Sigma\xstar}),
\end{equation}
where again $t_{0.025,\nu}$ and $t_{0.975,\nu}$ are the 0.025th and 0.975th quantiles of a $t$-distribution with $\nu$ degrees of freedom;
as shown in~\ref{app:tdistribution}, the $t$-distribution converges to the normal distribution as $n\rightarrow\infty$, so that $t_{0.025,\nu}$ and $t_{0.975,\nu}$ are approximately -1.96 and 1.96 for large values of $\nu$, respectively.
The variance term can be rewritten as
\begin{equation}
\label{quadform}
\xstar'\Sigma\xstar=s^2\left(\frac{1}{n}+\frac{(x_*-\bar{x})^2}{\sum_{i=1}^n(x_i-\bar{x})^2}\right),
\end{equation}
which increases with the distance from $x_*$ to $\bar{x}$ and is expected to decrease with increasing $n$ for shock compression experiments where the particle velocities are bounded.
Hence, credible intervals for the mean shock wave velocity are narrowest at the mean particle velocity and tend to shrink as the number of measurements increases.

Uncertainty in future shock wave velocity measurements is quantified with the posterior predictive distribution.
If the new shock wave velocity measurement is denoted $\tilde{Y}$ and $\tilde{y}$ is a particular value that $\tilde{Y}$ can take on, then the posterior predictive distribution is $p(\tilde{y}|Y)$, which is computed from the measurement distribution and posterior distribution as
\begin{equation}
\label{ppd_density}
p(\tilde{y}|Y)=\int p(\tilde{y}|\beta,\sigma^2)p(\beta,\sigma^2|Y)\,d\beta\,d\sigma^2.
\end{equation}
As derived in~\ref{app:ppd}, the posterior predictive distribution at a particle velocity $x_*$ is
\begin{equation}
\label{post_pred_dist}
\tilde{Y}|Y\sim t_1(\xstar'\hat{\beta},s^2+\xstar'\Sigma\xstar,\nu).
\end{equation}
The mean of this distribution is the least squares prediction, $\xstar'\hat{\beta}$, and the scale parameter is the sum of the sample variance of the residuals, $s^2$, and the scale parameter of the distribution of $\us|Y$ given in~\eqref{post_dist_us}.
Hence, the variance of the posterior predictive distribution accounts for uncertainty from the new measurement error and from the posterior distribution of the shock wave velocity.
As the number of measurements increases, the variance of the posterior predictive distribution converges to the variance of the measurement error, $\sigma^2$, since $s^2$ converges to $\sigma^2$, and $\xstar'\Sigma\xstar$ converges to zero from~\eqref{quadform}.

\begin{figure}[h!]
  \centering
  \begin{subfigure}[b]{0.32\textwidth}
    \centering
    \includegraphics[width=\textwidth]{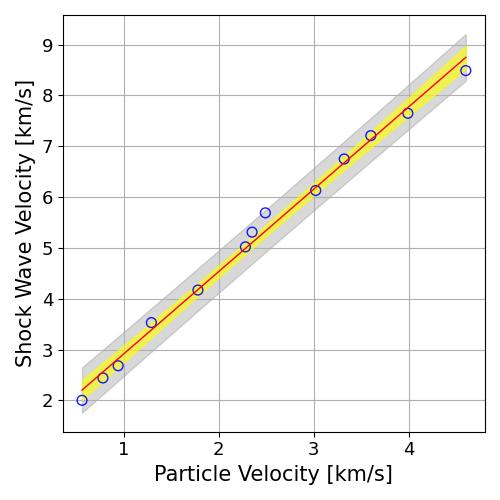}
    \caption{Argon}
  \end{subfigure}
  \begin{subfigure}[b]{0.32\textwidth}
      \centering
      \includegraphics[width=\textwidth]{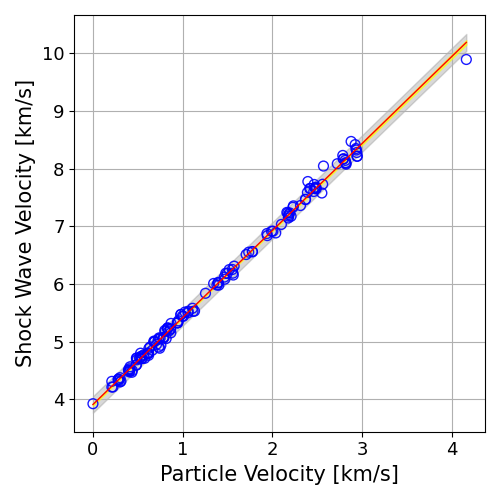}
      \caption{Copper}
  \end{subfigure}
  \begin{subfigure}[b]{0.32\textwidth}
    \centering
    \includegraphics[width=\textwidth]{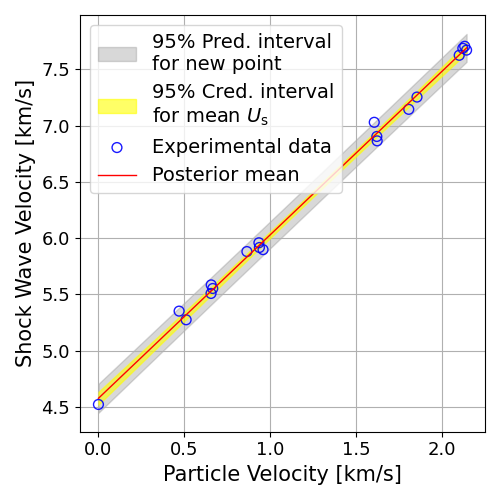}
    \caption{Nickel}
  \end{subfigure}
  \caption{Credible intervals for mean shock wave velocities and prediction intervals for future shock wave velocity measurements, obtained from the posterior distributions of $C_0$, $S$, and $\sigma^2$.}
  \label{fig:pred_ints}
\end{figure}

Figure~\ref{fig:pred_ints} shows 95\% credible intervals for the mean shock wave velocity and 95\% prediction intervals for future shock wave velocity measurements.
The latter are computed as quantiles of the posterior predictive distribution from~\eqref{post_pred_dist} as
\begin{equation}
(\xstar'\hat{\beta}+t_{0.025,\nu}\sqrt{s^2+\xstar'\Sigma\xstar},\xstar'\hat{\beta}+t_{0.975,\nu}\sqrt{s^2+\xstar'\Sigma\xstar}),
\end{equation}
which differs from the credible interval given in~\eqref{credint_Us} by an extra $s^2$ in the variance term to account for measurement error.
Two observations from these plots are readily apparent.
First, the credible interval width decreases with the sample size, as expected from the previous discussion about the variance of the posterior predictive distribution decreasing as the number of measurements increases.
This is evident in the figure since the credible intervals for copper are barely visible, whereas the credible intervals for argon are easy to see.
Second, the prediction intervals tend to contain the shock wave velocity measurements, except for copper, which has several points falling outside the prediction bands.
This is not surprising since only 95\% of future measurements are expected to be contained in the prediction intervals, and furthermore prediction intervals are designed to contain future measurements with some specified probability, not current measurements.
Assessing prediction interval coverage on the data used to construct the posterior distribution is mainly meant as a heuristic to detect model fitting issues.

\begin{figure}[h!]
  \centering
  \begin{subfigure}[b]{0.32\textwidth}
    \centering
    \includegraphics[width=\textwidth]{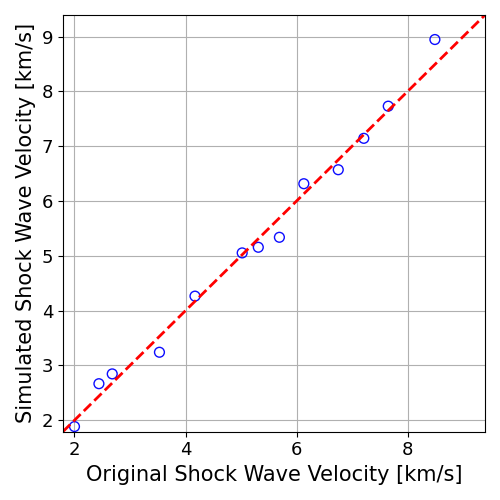}
    \caption{Argon}
  \end{subfigure}
  \begin{subfigure}[b]{0.32\textwidth}
      \centering
      \includegraphics[width=\textwidth]{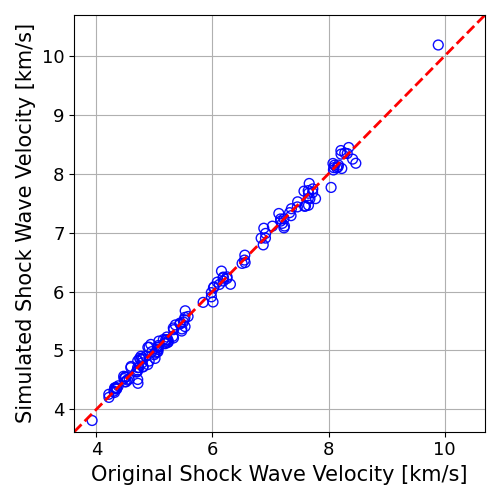}
      \caption{Copper}
  \end{subfigure}
  \begin{subfigure}[b]{0.32\textwidth}
    \centering
    \includegraphics[width=\textwidth]{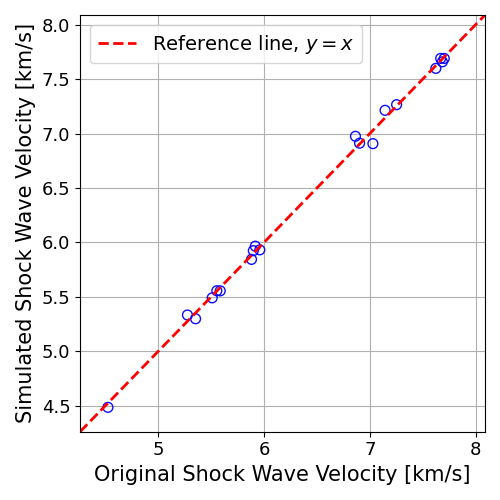}
    \caption{Nickel}
  \end{subfigure}
  \caption{Posterior predictive checks comparing the actual shock wave velocity measurements to measurements simulated from the posterior predictive distribution.}
  \label{fig:post_pred_check}
\end{figure}

To validate the fitted model, we simulate shock wave velocity measurements from the posterior predictive distribution and plot them against the actual shock wave velocity measurements.
This is called a posterior predictive check and is discussed more generally in~\citet[Sec. 6.3]{gelman2020}.
Figure~\ref{fig:post_pred_check} shows the results.
We see that the simulated and original measurements are close and differ by random error, which is expected since the original measurements had $R^2$ values close to one and both simulated and real measurements are generated with measurement error.
Deviation of the points from the red reference line, such as the points being systematically above or below the line, would suggest issues with the assumed linear model.
Hence, these plots do not suggest a serious issue with the fitted model, such as missing a systematic bias or mis-specifying the error model.
Note, however, that care needs to be taken when performing posterior predictive checks since different simulated datasets will produce different results, so it is advisable to perform posterior predictive checks with multiple sets of simulated data.
Also note that there are many ways to compare real and simulated data, and that we are showing a single visual approach.
Additional comparison methods that could be used here include plotting the residuals between the real and simulated shock wave velocities versus particle velocity, or plotting the quantiles of the real data against the quantiles of the simulated data in a quantile-quantile (QQ) plot~\citep{marden2004}.
The latter plot is created by sorting both the real and simulated shock wave velocities, and then plotting the two sorted datasets against each other.
Since the two datasets are sorted independently, points in the QQ plot may not have the same particle velocity, but the shock wave velocities should be close if the two distributions are similar.
Example QQ plots for the Swegle-Grady power law can be found in~\cite{gurrutxaga2025}.

\section{Comparison with Linear Regression and Bootstrapping}
\label{sec:comparison}

Two alternatives to the Bayesian method discussed here are linear regression and bootstrapping.
This section describes these methods and compares them to Bayesian linear regression.
Overall, our viewpoint is not that one method is preferred over another, but rather that they have different assumptions and interpretations, which this section aims to make clear.

The linear regression model for the shock wave velocities is given by~\eqref{obsmodel}, but in this case the $\beta$ and $\sigma^2$ parameters are treated as constants instead of random variables.
Consequently, there is no prior distribution on $\beta$ and $\sigma^2$ in non-Bayesian linear regression.
The maximum likelihood estimate (MLE) of $\beta$ is obtained by maximizing the likelihood given in~\eqref{likelihood}, which results in the estimate $\hat{\beta}$ given by~\eqref{beta_hat}.
The MLE is normally distributed with mean $\beta$ and covariance matrix $\sigma^2(X'X)^{-1}$, written
\begin{equation}
\hat{\beta}\sim N(\beta,\sigma^2(X'X)^{-1}).
\end{equation}
In practice, $\sigma^2$ is not known and so is estimated with the sample variance, $s^2$, which leads to confidence intervals for $\beta$ that coincide with the credible intervals obtained in the Bayesian framework.
However, confidence and credible intervals have different interpretations.
In non-Bayesian linear regression, $C_0$ is treated as a fixed but unknown parameter, and a 95\% confidence interval for $C_0$ is constructed such that, under repeated sampling, approximately 95\% of such intervals would contain the true value of $C_0$.
The same interpretation applies to the slope parameter $S$.
In contrast, with Bayesian linear regression, $\beta$ is treated as a random vector, and a 95\% credible interval for an element of $\beta$ is interpreted as an interval that, conditional on the data, contains that element with probability 0.95.
These differing interpretations should be considered when choosing to summarize uncertainty with confidence or credible intervals.
Furthermore, the intervals will not be equal if other prior distributions are placed on the model parameters.

Bootstrapping is another approach to inference that relies on creating new datasets that have similar statistical properties to the original dataset.
The approach we take, called non-parametric bootstrapping, involves creating synthetic datasets by sampling the original shock wave-particle velocity measurement pairs with replacement $n$ times.
A $\hat{\beta}$ is computed for each synthetic dataset using~\eqref{beta_hat}, which results in a sampling distribution of estimates of $C_0$ and $S$, referred to as a bootstrap distribution.
This method of bootstrapping is called non-parametric since the new datasets are not generated from a Gaussian parametric model.
In contrast, a parametric bootstrap distribution of $\hat{\beta}$ is generated by simulating new datasets from the Gaussian model fit to the original data, and then computing $\hat{\beta}$ for each new dataset.
Other bootstrap methods have also been developed, such as the Bayesian bootstrap~\citep{rubin1981}, but we do not provide details here on these alternatives.
More background on bootstrapping can be found in~\cite{hesterberg2011} or~\cite{diciccio1996}, for example.

\begin{figure}[h!]
  \centering
  \begin{subfigure}[b]{0.32\textwidth}
    \centering
    \includegraphics[width=\textwidth]{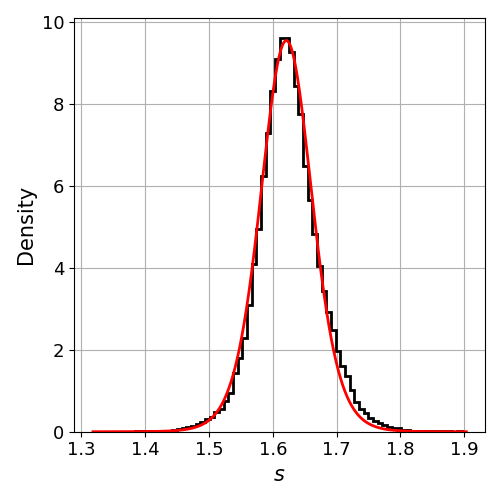}
    \caption{Argon}
  \end{subfigure}
  \begin{subfigure}[b]{0.32\textwidth}
      \centering
      \includegraphics[width=\textwidth]{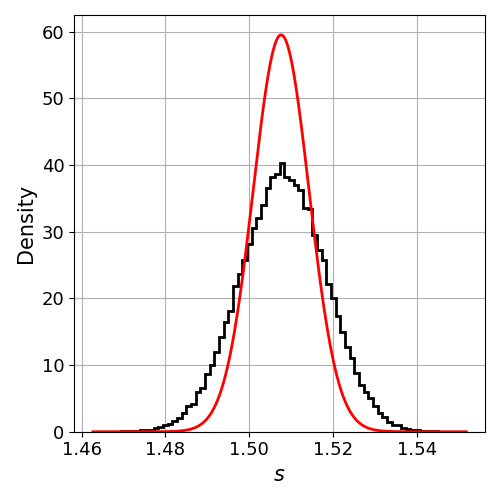}
      \caption{Copper}
  \end{subfigure}
  \begin{subfigure}[b]{0.32\textwidth}
    \centering
    \includegraphics[width=\textwidth]{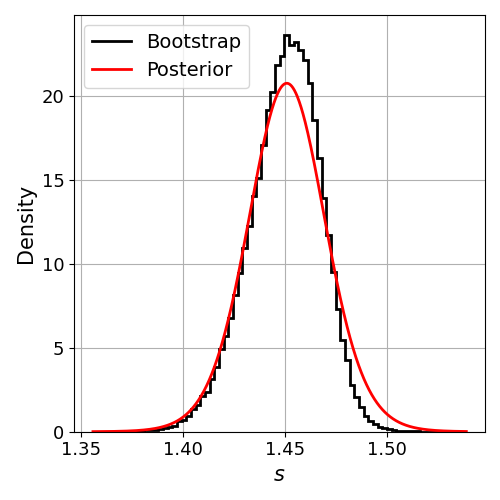}
    \caption{Nickel}
  \end{subfigure}
  \caption{Bootstrap distributions of $\hat{S}$ and posterior distributions of $S$.}
  \label{fig:comparison}
\end{figure}

Figure~\ref{fig:comparison} compares the bootstrap distribution of $\hat{S}$ with the posterior distribution of $S$ for the argon, copper, and nickel datasets.
Note that the bootstrap distributions are not necessarily centered around $\hat{S}$ and can be asymmetric, as seen for the nickel dataset.
In contrast, the posterior distributions have mean $\hat{S}$ and are symmetric since they are $t$-distributions.
The bootstrap distribution of $\hat{S}$ also has a visibly larger variance than the posterior distribution of $S$ for copper, though the variances appear much closer for argon and nickel.
The means, standard deviations, and percentile confidence intervals for the bootstrap distributions are given in Table~\ref{tab:bootstrap_regression_coefs};
the percentile confidence intervals are computed by taking the 2.5th and 97.5th percentiles of the bootstrap samples of $\hat{C}_0$ and $\hat{S}$.
In general, the bootstrap statistics agree up to the second decimal with the posterior statistics given in Table~\ref{tab:posterior_regression_coefs}, with some larger differences, mainly for the argon dataset, which is the smallest dataset.
Figure~\ref{fig:bootstrap_ints} in~\ref{app:bootstrap_intervals} shows bootstrap confidence intervals on the mean shock wave velocity and prediction intervals on new shock wave measurements.

\begin{table}[ht]
\caption{Bootstrap means, standard deviations, and 95\% percentile confidence intervals for $C_0$ and $S$.}
\label{tab:bootstrap_regression_coefs}
\centering
\begin{tabular}{|l|l|c|c|c|c|}
\hline
\textbf{Material} & \textbf{Parameter} & \makecell[c]{\textbf{Bootstrap}\\\textbf{Mean}} & \makecell[c]{\textbf{Bootstrap Standard}\\\textbf{Deviation}} & \makecell[c]{\textbf{95\% Percentile}\\\textbf{Confidence Interval}} & \textbf{Units} \\
\hline
\multirow{2}{*}{Argon}  & $C_0$ & $1.297$ & 0.122 & (1.106, 1.588) & \unit{km/s} \\
                        & $S$   & $1.625$ & 0.049 & (1.530, 1.727) & $\cdot$ \\
\hline
\multirow{2}{*}{Copper} & $C_0$ & $3.912$ & 0.012 & (3.890, 3.936) & \unit{km/s} \\
                        & $S$   & $1.508$ & 0.010 & (1.488, 1.528) & $\cdot$ \\
\hline
\multirow{2}{*}{Nickel} & $C_0$ & $4.579$ & 0.026 & (4.533, 4.635) & \unit{km/s}\\
                        & $S$   & $1.450$ & 0.017 & (1.413, 1.481) & $\cdot$ \\
\hline
\end{tabular}
\end{table}

A key difference between the form of bootstrapping employed here and the Bayesian approach is that the bootstrap distribution of $\hat{\beta}$ is obtained by resampling the original dataset, whereas the posterior distribution of $\beta$ is obtained holding the dataset fixed.
Due to the data resampling process, the $\hat{\beta}$ resulting from the bootstrap procedure are not $t$-distributed, so that the bootstrap confidence and prediction intervals are not computed from the $t$-distribution.
However, the posterior distribution summary statistics in Table~\ref{tab:posterior_regression_coefs} and bootstrap distribution summary statistics in Table~\ref{tab:bootstrap_regression_coefs} often agree to one or two decimal places.

\begin{figure}[h!]
  \centering
  \begin{subfigure}[b]{0.32\textwidth}
    \centering
    \includegraphics[width=\textwidth]{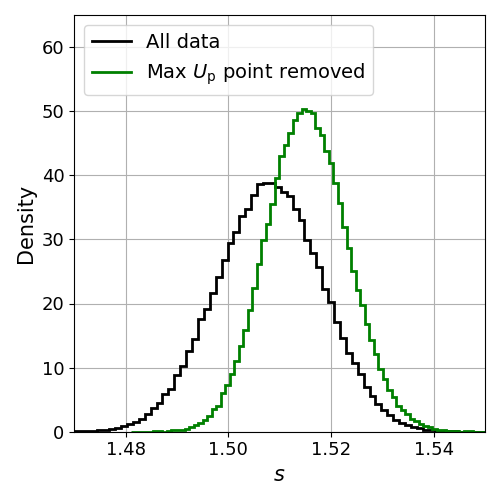}
    \caption{Bootstrap Distribution}
  \end{subfigure}
  \begin{subfigure}[b]{0.32\textwidth}
    \centering
    \includegraphics[width=\textwidth]{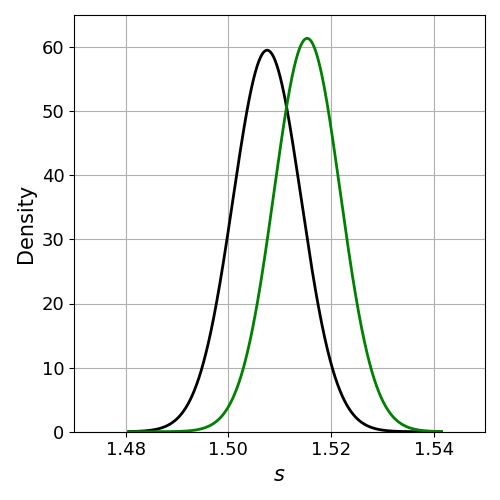}
    \caption{Posterior Distribution}
  \end{subfigure}
  \caption{Bootstrap distributions of $\hat{S}$ and posterior distributions of $S$ for the complete copper dataset and for the dataset with the point that has the largest particle velocity removed.}
  \label{fig:outlier}
\end{figure}

Bootstrap distributions can also be more sensitive to individual points than posterior distributions.
Figure~\ref{fig:outlier} shows the bootstrap distributions and posterior distributions of $S$ for the copper dataset with and without the point that has the largest particle velocity.
This is the point that has a particle velocity greater than 4 \unit{km/s} in Fig.~\ref{fig:cu}.
Removing this point increases the mean of the bootstrap and posterior distributions by about 0.007, but decreases the variance of the bootstrap distribution.
The variance of the bootstrap distribution is larger when the point is included in the original dataset because the point may still be excluded from some of the resampled datasets due to random chance.
In contrast, the variance of the posterior distribution appears similar whether the point is included or not because the posterior distribution quantifies the uncertainty in $S$ while holding the dataset fixed.
These results remain unchanged if the number of bootstrap datasets is doubled to 200,000.
Furthermore, if any other point in the copper dataset is removed, then the most that the estimate of $S$ changes by is 0.003, demonstrating that the previously removed point is influential and that the least squares estimate of $S$ is robust to removal of other points.
The argon and nickel datasets also have points that influence the posterior distributions of $\beta$.
Removing the point from the argon dataset that has the largest particle velocity changes the posterior mean of $S$ by 0.043, which is not surprising considering the dataset only has 13 points.
Similarly, excluding the sound speed experiment point from the nickel dataset changes the posterior mean of $S$ by 0.011.

A point to emphasize with non-parametric bootstrapping is that due to the resampling process, a $\hat{\beta}$ computed from a new dataset is likely to not be informed by all the measurements in the original dataset.
For example, if a dataset consists of $n=10$ shock wave and particle velocity measurements, then each bootstrap sample of $\hat{\beta}$ will be based on a random resampling, with replacement, of those ten measurements, and therefore will exclude a specified measurement with probability $0.9^{10}\approx 0.35$.
This outcome is expected with bootstrapping, and is offset by the specified measurement being included in other bootstrap samples of $\hat{\beta}$.
In contrast, in the Bayesian approach, each sample of $\beta$ obtained from sampling the posterior distribution of $\beta$ is informed by each measurement in the original dataset.

One advantage of the bootstrap for this application is that it does not make any assumptions about the distribution of the measurement errors or require a prior distribution.
In contrast, the Bayesian approach assumed independent Gaussian-distributed measurement errors and a non-informative prior distribution on the model parameters.
If the distributional assumptions made in the Bayesian analysis lead to a model that is inconsistent with the data, then the sampled Hugoniot curves in the pressure-volume plane will also be inconsistent with the data.
Given the advantages and disadvantages of the two methods, there is no clear decision rule for deciding whether to analyze shock compression data using a bootstrapping or Bayesian approach.
In the EOS context, we suggest validating the uncertainty quantification from either method by comparing predictions from the sampled Hugoniot curves to observables, and also considering the interpretation of the uncertainty quantification produced by each method.
If there is prior information about the linear model parameters or the parameters can be considered to be random variables, then a Bayesian analysis may be a better approach.
Alternatively, if there is no prior information about the parameters or if it is preferred to treat the parameters as fixed but unknown constants, then bootstrapping is likely to be more appealing.

It is also worth noting that both bootstrapping and the Bayesian approach presented here are computationally inexpensive.
The Bayesian analysis requires sampling from a bivariate $t$-distribution, which was shown in Sec.~\ref{sec:uncprop} to require sampling from a bivariate normal and chi-square distribution.
Bootstrapping, in comparison, requires resampling the original dataset multiple times and fitting a least squares line to each of the new datasets.
For the three datasets considered in this paper, this resampling and estimation procedure was performed 100,000 times, and the total time for performing the analysis on a standard laptop was approximately 25 seconds.

\section{Informative Prior Analysis}
\label{sec:informative_prior}

In some cases, there is prior information about $\beta$ and $\sigma^2$ that can be used to construct a prior distribution.
For example, if $C_0$ and $S$ are believed to be independent Gaussian random variables with a certain mean, then their prior distribution can be set to a bivariate normal distribution with a correlation of zero.
Additionally, if the beliefs are not strong, then the prior variances on the parameters can be made large.
In this section, we discuss and provide results for an analysis with an informative prior distribution, but leave mathematical details and derivations to~\ref{app:informative_prior}. 

A normal-inverse-gamma distribution is a common choice of prior distribution when prior knowledge is available because it leads to a normal-inverse-gamma posterior distribution, a property referred to as conjugacy.
This distribution was encountered in Sec.~\ref{sec:blr} when the posterior distribution for $\beta$ and $\sigma^2$ under a non-informative prior distribution was identified as a normal-inverse-gamma distribution;
specifically, the distribution of $\beta$ given $\sigma^2$ and $Y$ was Gaussian by~\eqref{posterior_beta_given_sigmasq_Y} and the distribution of $\sigma^2$ given $Y$ was inverse gamma by~\eqref{posterior_sigmasq}.
Similar results hold for the informative prior case, with the corresponding distributions given in~\eqref{posterior_beta_informative_prior} and~\eqref{posterior_sigmasq_informative_prior}, respectively.
Furthermore, the marginal posterior distribution of $\beta$ under an informative prior distribution is still a bivariate $t$-distribution, and is given in~\eqref{marginal_posterior_beta_informative_prior}.
In fact, the posterior distributions for the non-informative and informative prior distribution cases are both normal-inverse-gamma distributions because the non-informative prior distribution can be viewed as a limit of the informative prior distribution.
Specifically, an informative, normal-inverse-gamma prior distribution converges to the non-informative prior distribution in~\eqref{improper_prior} as the parameters of the distribution converge to limiting values given in~\ref{app:informative_prior}.

\begin{figure}[h!]
  \centering
  \begin{subfigure}[b]{0.32\textwidth}
    \centering
    \includegraphics[width=\textwidth]{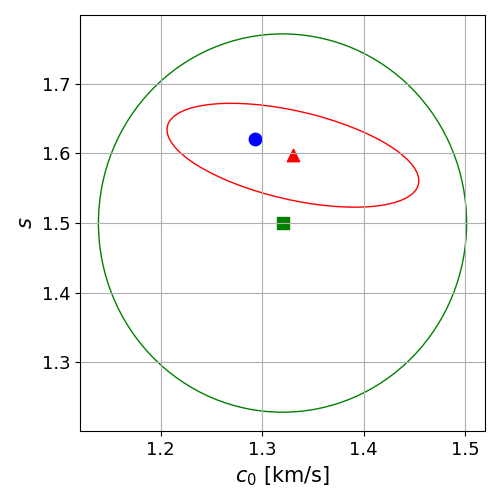}
    \caption{Argon}
  \end{subfigure}
  \begin{subfigure}[b]{0.32\textwidth}
      \centering
      \includegraphics[width=\textwidth]{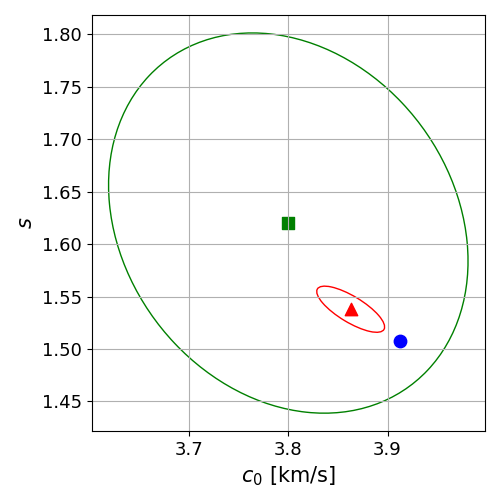}
      \caption{Copper}
  \end{subfigure}
  \begin{subfigure}[b]{0.32\textwidth}
    \centering
    \includegraphics[width=\textwidth]{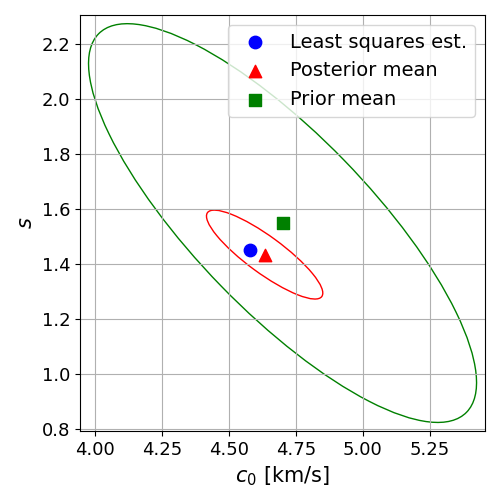}
    \caption{Nickel}
  \end{subfigure}
  \caption{Posterior $t$-distributions for $\beta=(C_0,S)'$ under informative prior distributions. The green ellipses are 95\% probability regions for $\beta$ under the prior distribution and the red ellipses are 95\% credible regions for $\beta$ under the posterior distribution. Each posterior mean is a weighted average of the least squares estimate and the prior mean.}
  \label{fig:posterior_beta_informative}
\end{figure}

To illustrate the effect of an informative prior distribution on the posterior distribution, Fig.~\ref{fig:posterior_beta_informative} shows the marginal posterior distributions of $\beta$ assuming an informative, normal-inverse-gamma prior distribution on $\beta$ and $\sigma^2$.
The red ellipses are 95\% credible regions from the posterior distribution, and the green ellipses are the analogous ellipses containing 95\% of the mass of the prior probability distribution.
Hence, the reduction in width of the red ellipses relative to the green ellipses indicates that the variances of the posterior distribution are smaller than the variances of the prior distribution, showing how the data reduce uncertainty in $\beta$.
Additionally, as suggested by the figure and shown mathematically in~\ref{app:informative_prior}, the posterior mean is a weighted average of the least squares estimate, $\hat{\beta}$, and the prior mean of $\beta$.
The sample size, $n$, and the prior variances on $\beta$ determine how much the posterior mean is pulled away from the least squares estimate and toward the prior mean, with the effect of the prior mean diminishing as the sample size and prior variances increase.
For the three materials, the posterior means are closer to the least squares estimates than to the prior means, reflecting that the data influence the posterior distributions more than the chosen prior distributions.

To generate Fig.~\ref{fig:posterior_beta_informative}, the prior correlations for argon, copper, and nickel were set to 0, -0.2, and -0.8, respectively, which explains why the green ellipses appear increasingly diagonally slanted for these materials.
When an informative prior is used, the posterior correlation is influenced by both the prior correlation and the experimental data, whereas the posterior correlation is not influenced by the prior correlation for the non-informative prior considered earlier in Sec.~\ref{sec:blr}.
The full set of parameters used to generate Fig.~\ref{fig:posterior_beta_informative} is given in Table~\ref{tab:informative_prior_hyperparameters}.

\section{Discussion}
\label{sec:discussion}

So far, our Bayesian analysis has focused on computing and interpreting the posterior distribution under non-informative and informative priors, validating the fitted model using posterior predictive checks, and assessing the sensitivity of the posterior distribution to a possible outlier.
This section briefly covers additional topics, including prior selection and model inadequacy issues, that should also be considered in an analysis of shock wave-particle velocity data.

In general, posterior distributions can be sampled from approximately or computed numerically when they cannot be sampled directly or are not analytically available.
Approximate sampling for our regression model would be necessary if the prior distribution was not a normal-inverse-gamma distribution, for example.
One approach to obtaining approximate samples from a posterior distribution is to use an iterative, probabilistic algorithm called Markov Chain Monte Carlo (MCMC)~\citep[Ch. 11]{gelman1995}.
The \verb|emcee| package in Python is a popular choice for implementing MCMC~\citep{foreman2013}, though care should be taken if the number of parameters to sample is large~\citep{huijser2022}.
An MCMC algorithm called a Gibbs sampler can also be used to draw samples from the posterior distribution of the linear model parameters, as described in~\citet[Sec. 8.3]{chib2001}.
Note that these MCMC algorithms produce a sequence of correlated samples from the posterior distribution, whereas the direct sampling methods presented in this paper produce independent samples.
Alternatively, since the linear regression model examined in this paper only has three parameters, the posterior distribution can be evaluated on a three-dimensional grid.
That is, the prior distribution is multiplied by the likelihood for each grid cell to obtain the non-normalized posterior distribution, and then these values are normalized to sum to one to approximate the actual posterior distribution.

This discussion highlights the fact that the Bayesian paradigm offers great flexibility in terms of modeling choices and inferential procedures.
Some analysts will view this flexibility as a strength since it allows them to examine a range of models and tailor the inferential approach to the model.
Other analysts may view this flexibility as a nuisance and prefer to stick to a general computational method that works for a broad class of Bayesian models.
Overall, using a more general approach than necessary is not problematic, as long as the same posterior distribution is computed or sampled.
For example, MCMC can be used to sample the posterior distribution of $(\beta,\sigma^2)$ for the Bayesian linear model presented here, and it will be relatively inefficient compared to sampling the actual normal-inverse-gamma posterior distribution directly, but both approaches target the same posterior distribution.

Several important topics were not covered in this tutorial but are now briefly mentioned for completeness:
\begin{enumerate}
\item In many scientific contexts, the model fit to the data is an approximation, and this approximation can introduce systematic error.
For example,~\cite{brynjarsdottir2014} gives an example of a model for work that always predicts less than the actual work.
A possible explanation for this systematic discrepancy is that the model to be calibrated is missing a friction term. 
This model discrepancy is then modeled using a Gaussian process, which is also inferred while performing Bayesian calibration.
The shock wave-particle velocity data considered here is linear, as indicated by the $R^2$ values close to one, so that this source of error did not require consideration.
However, other shock wave-particle velocity datasets in~\cite{marsh1980} are non-linear and could be analyzed with this Gaussian process model approach.
\item The particle velocity measurements were assumed to be known exactly, but it is possible to account for particle velocity measurement error in the Bayesian modeling and calibration framework.
The resulting posterior distribution is no longer available in closed form, but it can be sampled from using MCMC;
see~\citet[Ch. 9]{carroll2006} for details and examples.
In general, the effect of observing particle velocities with measurement error will be to flatten the regression line, which is called attenuation~\citep[Ch. 3]{carroll2006}.
\cite{celliers2005} provide a non-Bayesian analysis of systematic uncertainty in impedance matching experiments.
\item A common criticism of Bayesian analysis is that it is subjective in the sense that the analyst chooses the prior distribution, and different results can be obtained with different prior distributions.
Nonetheless, being able to select a prior distribution is helpful when there is limited data or when there is subject matter expertise that can be brought into the analysis.
The criticism can also be mitigated by performing a prior sensitivity analysis, as discussed in a different context in~\citet[Sec. 5.9]{gelman2020}.
For example, as shown in Fig.~\ref{fig:posterior_beta_informative}, the posterior mean of $\beta$ under an informative prior distribution on $\beta$ can be far from the prior mean but close to the least squares estimate.
This indicates that the posterior distributions shown in the figure are driven by the data and not the prior distributions.
In general, the difficulty with assuming a normal prior distribution on $\beta$ is specifying the mean of the distribution, which we effectively avoided by choosing a non-informative prior distribution.
\item If a shock wave-particle velocity dataset includes a phase change, then the relationship between the velocities may be piecewise linear with the change in slope occurring at the phase change.
Bayesian methods can still be used to calibrate this model by including both slopes and the location of the change in slope in the set of unknown parameters, and then sampling the posterior distribution of these parameters using MCMC.
See Fig.~16 in~\cite{duvall1977} for an example of shock wave-particle velocity data that suggests a phase transition due to a deviation from linearity at large shock wave velocities.
\item This paper assumed that shock wave velocity was linearly related to particle velocity, but quadratic and cubic models can also be found in the literature.
For example,~\cite{root2019} fit a model that is cubic in particle velocity,
\begin{equation}
\us=C_0 + C_1\up + C_2\up^2 + C_3\up^3,
\end{equation}
which fits into our modeling framework with $\beta=(C_0,C_1,C_2,C_3)'$ and $\nu=n-4$;
see their Table 4 for estimates of these model parameters for fused silica.
\end{enumerate}
We leave the implementation and further investigation of these points as future work or for interested readers.

While the previous discussion highlights several problems related to linear EOS models, the literature review in Sec.~\ref{sec:introduction} suggests that the broader shock and dynamic compression community is moving toward more complex data and computationally expensive physics models, for which linear models are inadequate.
Machine learning models, such as neural networks, will likely be pursued instead, though these models often do not provide uncertainty quantification as naturally as linear or Gaussian process models.
A challenge will therefore be to incorporate these machine learning models into Bayesian workflows, with Bayesian neural networks~\citep{jospin2022} and generative artificial intelligence (AI) models~\citep{polson2025} offering two approaches.
We emphasize that the same considerations and diagnostics we discussed for a linear model, such as prior sensitivity and posterior predictive checks, are still required for these more complex models.

\section{Conclusion}
\label{sec:conclusion}

This paper showed how to analyze linear shock compression data in a Bayesian framework.
Specifically, under certain standard assumptions, the posterior distribution of the linear model parameters is a bivariate $t$-distribution that can be sampled directly.
These samples can then be propagated through the Rankine-Hugoniot equations to obtain Hugoniot curves in the pressure-volume plane, or alternatively the posterior distribution can be propagated analytically through the shock wave-particle velocity model to quantify the uncertainty in future shock wave measurements.
The method was shown to produce parameter distributions that are less sensitive to the removal of a point with large particle velocity than bootstrapping, though use of either method should also take into account the interpretation of the distributions they produce.

Overall, evaluating the posterior distribution is straightforward and sampling from the distribution is simple, but this perspective on shock compression data also appears to have been overlooked in the literature.
For these reasons, this tutorial both fills a gap in the literature and serves as a useful starting point for practitioners looking to gain familiarity with Bayesian analysis.
Additionally, it provides a benchmark for more advanced Bayesian linear model analyses, such as accounting for errors in the particle velocity measurements or relaxing the assumption of independent errors on the shock wave velocity measurements.
If either of these more complex models is implemented, its results can be compared against those from the simpler model presented here.

\newpage

\section*{Acknowledgements}

The authors thank two anonymous reviewers for their comments and suggestions that led to improvements in the paper.

\begin{quote}
This document was prepared as an account of work sponsored by an agency of the United States government.
Neither the United States government nor Lawrence Livermore National Security, LLC, nor any of their employees makes any warranty, expressed or implied, or assumes any legal liability or responsibility for the accuracy, completeness, or usefulness of any information, apparatus, product, or process disclosed, or represents that its use would not infringe privately owned rights.
Reference herein to any specific commercial product, process, or service by trade name, trademark, manufacturer, or otherwise does not necessarily constitute or imply its endorsement, recommendation, or favoring by the United States government or Lawrence Livermore National Security, LLC.
The views and opinions of authors expressed herein do not necessarily state or reflect those of the United States government or Lawrence Livermore National Security, LLC, and shall not be used for advertising or product endorsement purposes.
\end{quote}

\begin{quote}
Lawrence Livermore National Laboratory is operated by Lawrence Livermore National Security, LLC, for the U.S. Department of Energy, National Nuclear Security Administration under Contract DE-AC52-07NA27344.

\textbf{Release Numbers}: LLNL-JRNL-2014240, LA-UR-26-21559
\end{quote}

\section*{Author Declarations}

\subsection*{Conflict of Interest}

The authors have no conflicts to disclose.

\subsection*{Author Contributions}

\noindent
\textbf{Jason Bernstein:}
Conceptualization (equal);
Formal analysis (equal);
Investigation (equal);
Methodology (equal);
Software (lead);
Visualization (lead);
Writing~-~original draft (lead);
Writing~-~review \& editing (equal).
\textbf{Philip Myint:}
Conceptualization (equal);
Formal analysis (equal);
Investigation (equal);
Methodology (equal);
Writing~-~original draft (supporting);
Writing~-~review \& editing (equal).
\textbf{Beth Lindquist:}
Conceptualization (equal);
Formal analysis (equal);
Investigation (equal);
Methodology (equal);
Writing~-~original draft (supporting);
Writing~-~review \& editing (equal).
\textbf{Justin Brown:}
Conceptualization (equal);
Formal analysis (equal);
Investigation (equal);
Methodology (equal);
Writing~-~original draft (supporting);
Writing~-~review \& editing (equal).

\section*{Data Availability}

The data that support the findings of this study were originally published in~\cite{marsh1980} and can be downloaded from the public BALSCD repository (\repourl).

\appendix
\renewcommand{\theHequation}{\thesection.\arabic{equation}}

\renewcommand{\thesection}{Appendix \Alph{section}}
\makeatletter
\renewcommand{\@seccntformat}[1]{\csname the#1\endcsname:\space}
\makeatother

\renewcommand{\theequation}{\Alph{section}\arabic{equation}}

\section{Overview of Multivariate $t$-Distribution}
\label{app:tdistribution}
\setcounter{equation}{0}

As noted in Sec.~\ref{sec:blr}, the marginal posterior distribution of $\beta=(C_0,S)'$, or $p(\beta|Y)$, follows a multivariate $t$-distribution. 
A random vector $X$ is said to have a $p$-dimensional $t$-distribution with location parameter $\mu$, scale matrix $\Sigma$, and $\nu$ degrees of freedom if it has the probability density function
\begin{equation}
\label{tpdf}
f(x)=C(\nu,p,\Sigma)\left[1+\frac{1}{\nu}(x-\mu)'\Sigma^{-1}(x-\mu)\right]^{-(\nu+p)/2}
\end{equation}
for $x\in\mathbb{R}^p$ and
\begin{equation}
C(\nu,p,\Sigma)=\frac{\Gamma((\nu+p)/2)}{\Gamma(\nu/2)\nu^{p/2}\pi^{p/2}|\Sigma|^{1/2}},  
\end{equation}
which is notated
\begin{equation}
X\sim t_p(\mu,\Sigma,\nu).
\end{equation}
If $\nu>1$, then the mean of the distribution is $\mu$.
The normalizing constant, $C(\nu,p,\Sigma)$, ensures that $f(x)$ integrates to one over $\mathbb{R}^p$.
When $\nu=1$, the multivariate $t$-distribution is a multivariate Cauchy distribution.

\begin{figure}[ht]
    \centering
    \includegraphics[width=0.8\textwidth]{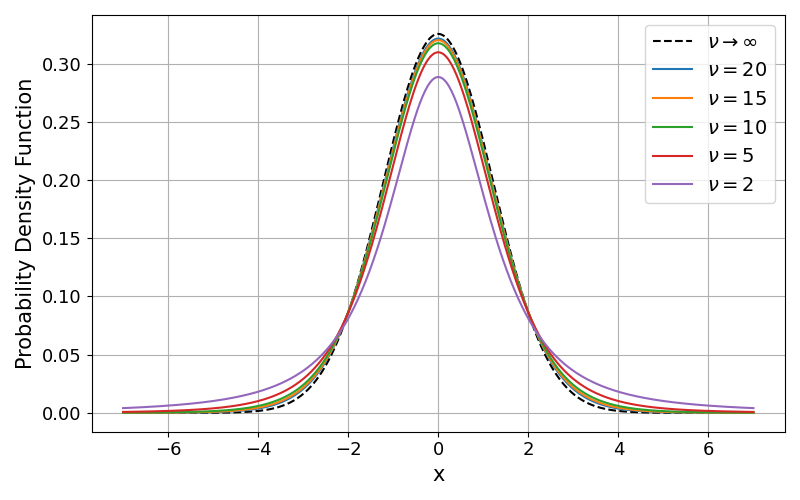}
    \caption{The $t$-distribution converges to a normal distribution as the degrees of freedom increase to infinity. For this plot, each $t$-distribution has dimension $p=1$, mean $\mu=0$, and scale parameter $\Sigma=1.5$. The dashed line is the density of a univariate normal distribution with mean $\mu=0$ and variance $\sigma^2=1.5$.}
    \label{fig:t_convergence}
\end{figure}

Figure~\ref{fig:t_convergence} illustrates convergence of the univariate $t$-distribution to the normal distribution as $\nu\rightarrow\infty$.
For $\nu\geq3$, we see that the $t$-distributions have greater variance than the limiting normal distribution since they have variance $\nu/(\nu-2)\Sigma$, though this difference becomes negligible as $\nu$ increases.
In practice, the univariate $t$-distribution is often approximated as a normal distribution when $\nu\geq30$.

We can also show that the multivariate $t$-distribution converges to the multivariate normal distribution as $\nu\rightarrow\infty$ in two steps.
First, it follows from~\citet[eqn. 6.1.46]{abramowitz1964} that
\begin{equation}
\lim_{\nu\rightarrow\infty} C(\nu,p,\Sigma)=\frac{1}{2^{p/2}\pi^{p/2}|\Sigma|^{1/2}}.
\end{equation}
Second, the limit of the second term in the probability density function in~\eqref{tpdf} is
\begin{equation}
\lim_{\nu\rightarrow\infty}\left[1+\frac{1}{\nu}(x-\mu)'\Sigma^{-1}(x-\mu)\right]^{-(\nu+p)/2}=\exp\left(-\frac{1}{2}(x-\mu)'\Sigma^{-1}(x-\mu)\right),
\end{equation}
which follows from the exponential function identity
\begin{equation}
\lim_{\nu\rightarrow\infty}\left(1+\frac{z}{\nu}\right)^{-\nu/2}=\exp\left(-\frac{z}{2}\right)
\end{equation}
with $z=(x-\mu)'\Sigma^{-1}(x-\mu)$.
Hence, combining the two limits, it follows that the density function of the multivariate $t$-distribution converges to the density function of a multivariate normal distribution with mean $\mu$ and covariance matrix $\Sigma$.
In the $p=2$ case considered in this paper, the numerator of $C(\nu,2,\Sigma)$ is $(\nu/2)\Gamma(\nu/2)$ by~\citet[eqn. 6.1.15]{abramowitz1964}, and so the normalizing constant is
\begin{equation}
C(\nu,2,\Sigma)=\frac{1}{2\pi|\Sigma|^{1/2}},
\end{equation}
which is independent of $\nu$.

An alternative proof of this convergence result is as follows.
Recall from~\eqref{tsample} that if $W\sim\chi^2_\nu$ and $Z\sim N(0,I)$ are independent, then $LZ/\sqrt{W/\nu}+\mu$ has a multivariate $t$-distribution with location parameter $\mu$ and scale matrix $\Sigma=LL'$.
Now, $W/\nu\rightarrow1$ in probability as $\nu\rightarrow\infty$ by the weak law of large numbers~\citep[Theorem 7.2.1]{resnick2019}, which implies that $\sqrt{W/\nu}\rightarrow1$ in probability as $\nu\rightarrow\infty$ by the continuous mapping theorem~\citep[Corollary 8.3.1]{resnick2019}.
Since $Z\sim N(0,I)$, it follows from Slutsky's theorem~\citep[Theorem 8.6.1]{resnick2019} that $Z/\sqrt{W/\nu}$ converges in distribution to a $N(0,I)$ distribution as $\nu\rightarrow\infty$, and so $LZ/\sqrt{W/\nu}+\mu$ converges in distribution to a $N(\mu,\Sigma)$ distribution.

\section{Joint Posterior Distribution of $(\beta,\sigma^2)$}
\label{app:joint_posterior}
\setcounter{equation}{0}

Under the non-informative prior distribution given in~\eqref{improper_prior}, the joint posterior distribution of $(\beta,\sigma^2)$ is derived as
\begin{align}
p(\beta,\sigma^2| Y)&\propto p(Y|\beta,\sigma^2)p(\beta,\sigma^2)\\
\label{tmp0}
&\propto\frac{1}{(\sigma^2)^{n/2}}\exp\left(-\frac{1}{2\sigma^2}(Y-X\beta)'(Y-X\beta)\right)\frac{1}{\sigma^2}\\
\label{tmp1}
&\propto\frac{1}{(\sigma^2)^{n/2}}\exp\left(-\frac{1}{2\sigma^2}(\beta'X'X\beta-2Y'X\beta+Y'Y)\right)\frac{1}{\sigma^2}\\
\label{tmp2}
&\propto\left[\frac{1}{|\sigma^2(X'X)^{-1}|^{1/2}}\exp\left(-\frac{1}{2\sigma^2}(\beta-\hat{\beta})'(X'X)(\beta-\hat{\beta})\right)\right]\left[\frac{1}{(\sigma^2)^{n/2}}\exp\left(-\frac{\nu s^2}{2\sigma^2}\right)\right]\\
\label{jointposterior}
&\propto p(\beta|\sigma^2,Y)p(\sigma^2|Y),
\end{align}
where the conditional distribution of $\beta$ given $\sigma^2$ and $Y$ is
\begin{equation}
p(\beta|\sigma^2,Y)\propto \frac{1}{|\sigma^2(X'X)^{-1}|^{1/2}}\exp\left(-\frac{1}{2\sigma^2}(\beta-\hat{\beta})'(X'X)(\beta-\hat{\beta})\right)
\end{equation}
and the marginal posterior distribution of $\sigma^2$ is
\begin{equation}
\label{pdf_sigmasq_given_Y}
p(\sigma^2|Y)\propto \frac{1}{(\sigma^2)^{n/2}}\exp\left(-\frac{\nu s^2}{2\sigma^2}\right).
\end{equation}
Note that the expression in~\eqref{tmp0} omits the $(2\pi)^{-n/2}$ term in~\eqref{likelihood} since it is a proportionality constant independent of $\beta$ and $\sigma^2$, and therefore is not needed to identify the joint posterior distribution of these parameters.
The expression in~\eqref{tmp1} is obtained by expanding the quadratic form, and the expression in~\eqref{tmp2} is obtained by completing a matrix square and using the identity
\begin{equation}
\nu s^2=Y'(I_n-X(X'X)^{-1}X')Y.
\end{equation}
The expression in~\eqref{jointposterior} implies that $\beta$, conditional on $\sigma^2$ and $Y$, has a multivariate normal distribution,
\begin{equation}
\label{beta_given_sigmasq_Y}
\beta|\sigma^2,Y\sim N(\hat{\beta},\sigma^2(X'X)^{-1}),
\end{equation}
and that the posterior distribution of $\sigma^2$ is an inverse gamma distribution,
\begin{equation}
\label{sigmasq_given_Y}
\sigma^2| Y\sim IG\left(\frac{\nu}{2},\frac{\nu s^2}{2}\right).
\end{equation}
As noted in Sec.~\ref{sec:blr}, the joint posterior distribution of $\beta$ and $\sigma^2$ is called the normal-inverse-gamma distribution due to this factorization.

\section{Marginal Posterior Distribution of $\sigma^2$}
\label{app:marginal_posterior_sigsq}
\setcounter{equation}{0}

\begin{figure}[h!]
  \centering
  \begin{subfigure}[b]{0.32\textwidth}
    \centering
    \includegraphics[width=\textwidth]{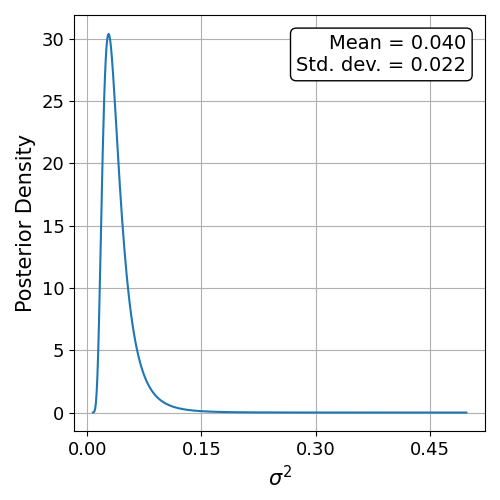}
    \caption{Argon}
  \end{subfigure}
  \begin{subfigure}[b]{0.32\textwidth}
      \centering
      \includegraphics[width=\textwidth]{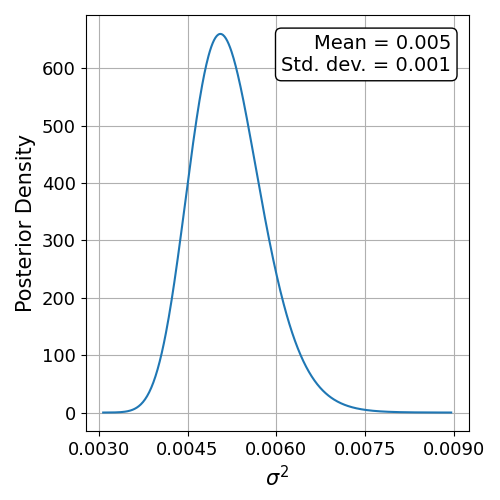}
      \caption{Copper}
  \end{subfigure}
  \begin{subfigure}[b]{0.32\textwidth}
    \centering
    \includegraphics[width=\textwidth]{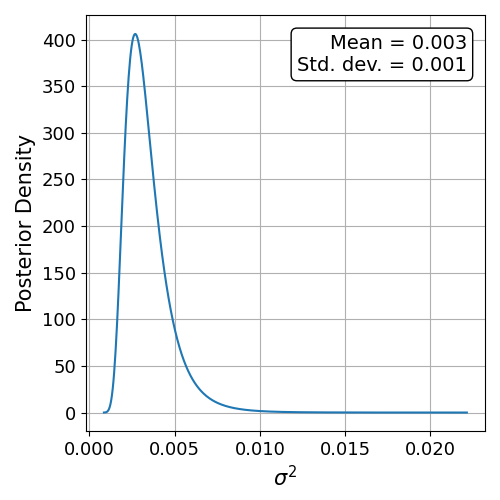}
    \caption{Nickel}
  \end{subfigure}
  \caption{Marginal posterior distributions of the variance of the measurement error, or $p(\sigma^2|Y)$.}
  \label{fig:posterior_sigma_sq}
\end{figure}

\begin{figure}[h!]
  \centering
  \begin{subfigure}[b]{0.32\textwidth}
    \centering
    \includegraphics[width=\textwidth]{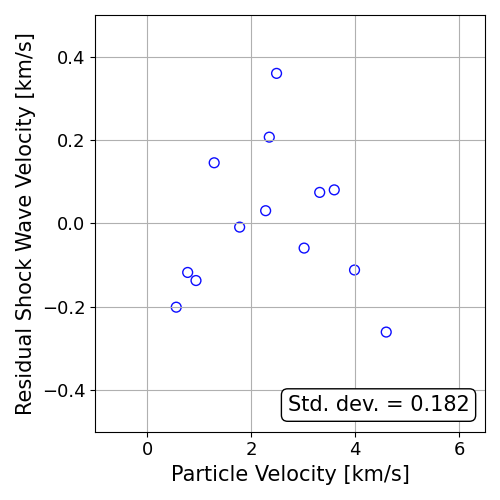}
    \caption{Argon}
  \end{subfigure}
  \begin{subfigure}[b]{0.32\textwidth}
      \centering
      \includegraphics[width=\textwidth]{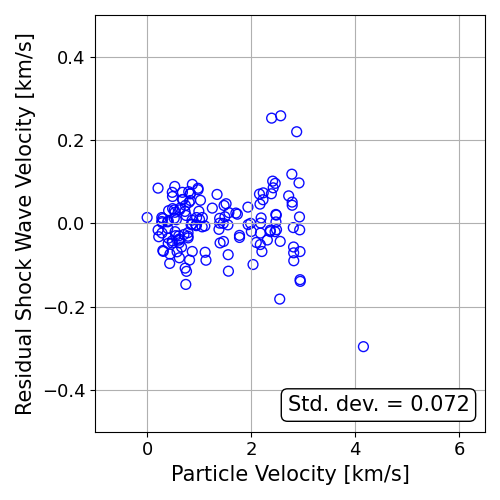}
      \caption{Copper}
  \end{subfigure}
  \begin{subfigure}[b]{0.32\textwidth}
    \centering
    \includegraphics[width=\textwidth]{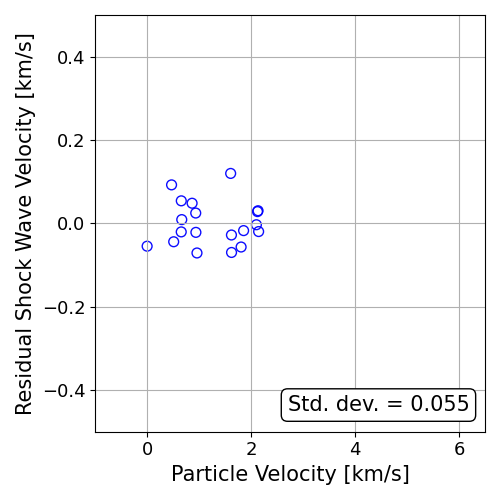}
    \caption{Nickel}
  \end{subfigure}
  \caption{Residual shock wave velocities from the least squares fits. The standard deviation of the residuals is the square root of the sample variance, $s^2$.}
  \label{fig:residuals}
\end{figure}

Figure~\ref{fig:posterior_sigma_sq} shows the posterior distributions of the measurement error variance parameter, $\sigma^2$, for the three datasets shown in Fig.~\ref{fig:data}.
Recall from~\eqref{posterior_sigmasq} that these are inverse gamma distributions with shape and scale parameters given by $a=(n-2)/2$ and $b=(n-2)s^2/2$, respectively.
The posterior mean of $\sigma^2$ is
\begin{equation}
\Expval{\sigma^2|Y}=\frac{b}{a-1},
\end{equation}
which is defined for $n>4$, and the posterior variance of $\sigma^2$ is
\begin{equation}
\Var{\sigma^2|Y}=\frac{b^2}{(a-1)^2(a-2)},
\end{equation}
which is defined for $n>6$.
The posterior mean and standard deviation are both shown in the upper right corner of the plots.
Note that the distributions are right skewed and that of these three datasets, the posterior mean and variance of $\sigma^2$ are greatest for argon.
This makes sense given that the residuals for argon appear to have greater spread, on average, than the residuals for copper or nickel, as shown in Fig.~\ref{fig:residuals}.
The skew does decrease as the number of measurements increases, with the distribution for copper appearing almost normal.

\section{Marginal Posterior Distribution of $\beta$}
\label{app:marginal_posterior_beta}
\setcounter{equation}{0}

The marginal posterior distribution of $\beta$ is obtained by integrating $\sigma^2$ out of the joint posterior distribution of $\beta$ and $\sigma^2$,
\begin{align}
p(\beta|Y)&=\int p(\beta,\sigma^2|Y)\,d\sigma^2\\
&=\int p(\beta|\sigma^2,Y)p(\sigma^2|Y)\,d\sigma^2\\
&\propto\int\frac{1}{(\sigma^2)^{n/2+1}}\exp\left(-\frac{1}{\sigma^2}\left[\frac{1}{2}\left(\nu s^2+(\beta-\hat{\beta})'(X'X)(\beta-\hat{\beta})\right)\right]\right)\,d\sigma^2\\
&\propto \left(\nu s^2+(\beta-\hat{\beta})'(X'X)(\beta-\hat{\beta})\right)^{-n/2}\\
&\propto \left(1+\frac{1}{\nu}(\beta-\hat{\beta})'\Sigma^{-1}(\beta-\hat{\beta})\right)^{-n/2}.
\end{align}
The last integral is computed using the observation that the integrand is proportional to the probability density function of an inverse gamma distribution with shape parameter $n/2$ and scale parameter given by the expression in the square brackets.
The last expression is the non-normalized density of a $t_2(\hat{\beta},\Sigma,\nu)$ distributed random variable, which implies that $\beta|Y\sim t_2(\hat{\beta},\Sigma,\nu)$.
Following~\ref{app:tdistribution}, the marginal posterior mean and covariance matrix of $\beta$ are $\hat{\beta}$ and $\nu/(\nu-2)\Sigma$, respectively.

The posterior mean of $\beta$ can also be computed with the law of total expectation as
\begin{align}
\Expval{\beta|Y}&=\Expval{\Expval{\beta|\sigma^2,Y}|Y}\\
&=\Expval{\hat{\beta}|Y}\\
&=\hat{\beta},
\end{align}
where the second equality follows from~\eqref{beta_given_sigmasq_Y}.
The posterior covariance matrix of $\beta$ can be computed with the law of total variance as
\begin{align}
\Var{\beta|Y}&=\Expval{\Var{\beta|\sigma^2,Y}|Y} + \Var{\Expval{\beta|\sigma^2,Y}|Y}\\
&=\Expval{\sigma^2(X'X)^{-1}|Y} + \Var{\hat{\beta}|Y}\\
&=(X'X)^{-1}\Expval{\sigma^2|Y}+(X'X)^{-1}X'\Var{Y|Y}X(X'X)^{-1}\\
&=(X'X)^{-1}\frac{(n-2)s^2/2}{(n-2)/2-1}\\
&=\frac{\nu}{\nu-2}\Sigma.
\end{align}
This derivation uses the fact that $\sigma^2|Y$ has an inverse gamma distribution given by~\eqref{posterior_sigmasq} and that $\Var{Y|Y}$ is an $n\times n$ zero matrix.

\section{Credible Region for $\beta$}
\label{app:credible_region_beta}
\setcounter{equation}{0}

Recall from Sec.~\ref{sec:blr} that a credible region for $\beta$ is a set of possible values of $\beta$, denoted $\mathcal{R}_\alpha$, such that $p(\beta\in\mathcal{R}_\alpha|Y)=1-\alpha$ for some $\alpha\in(0,1)$.
To derive a credible region for $\beta$, subtract $\hat{\beta}$ and then multiply both sides of~\eqref{tsample} by $L^{-1}$ to obtain
\begin{equation}
L^{-1}(\beta-\hat{\beta})=\frac{Z}{\sqrt{W/\nu}},
\end{equation}
or equivalently
\begin{equation}
\label{fdist_quad_form}
\frac{(\beta-\hat{\beta})'\Sigma^{-1}(\beta-\hat{\beta})}{2}=\frac{Z'Z/2}{W/\nu}.
\end{equation}
The right side of the last expression is the ratio of two independent chi-square random variables divided by their degrees of freedom, and therefore it has an $F$-distribution with 2 and $\nu$ degrees of freedom, denoted $F_{2,\nu}$.
Hence, a credible region is
\begin{equation}
\label{credible_region}
\mathcal{R}_\alpha=\{\beta|(\beta-\hat{\beta})'\Sigma^{-1}(\beta-\hat{\beta})\leq 2F_{1-\alpha,2,\nu}\},
\end{equation}
where $F_{1-\alpha,2,\nu}$ is the $(1-\alpha)$th quantile of an $F_{2,\nu}$ distribution.
The credible region is an ellipse whose semi-major and semi-minor axes have lengths $\sqrt{2\lambda_iF_{1-\alpha,2,\nu}}$ for $i=1$ and 2, where $\lambda_1$ and $\lambda_2$ are the eigenvalues of $\Sigma$.
To derive the semi-major and semi-minor axes, first diagonalize the scale matrix of the posterior distribution as
\begin{equation}
\Sigma=U\Lambda U',
\end{equation}
where $U$ is a matrix whose columns are the eigenvectors of $\Sigma$ and $\Lambda$ is a diagonal matrix whose diagonal elements are $\lambda_1$ and $\lambda_2$, and define $\eta=(\eta_1,\eta_2)'$ as
\begin{equation}
\eta=U'(\beta-\hat{\beta}).
\end{equation}
Plugging this expression for $\Sigma$ into the credible region and simplifying the resulting expression yields
\begin{equation}
\mathcal{R}_\alpha=\left\{\begin{pmatrix} \eta_1 \\ \eta_2 \end{pmatrix}
\,\middle|\,\frac{\eta_1^2}{2F_{1-\alpha,2,\nu}\lambda_1}+\frac{\eta_2^2}{2F_{1-\alpha,2,\nu}\lambda_2}\leq 1\right\},
\end{equation}
which is an ellipse with the claimed semi-major and semi-minor axes.
The red ellipses in Fig.~\ref{fig:posterior_beta} are boundaries of 95\% credible regions with $\alpha=0.05$.

\section{Credible Intervals for $C_0$ and $S$}
\label{app:credible_intervals_C0_S}
\setcounter{equation}{0}

To derive the credible interval for $C_0$ given in~\eqref{credint_C0}, first note that if the scale matrix $\Sigma$ is written as
\begin{equation}
\Sigma=\begin{pmatrix}
\sigma_1^2 & \rho\sigma_1\sigma_2 \\
\rho\sigma_1\sigma_2 & \sigma_2^2
\end{pmatrix},
\end{equation}
where $\rho\in(-1,1)$ is a correlation parameter, then the Cholesky decomposition of the scale matrix is
\begin{equation}
L=\begin{pmatrix}
\sigma_1 & 0 \\
\rho\sigma_2 & \sigma_2\sqrt{1-\rho^2}
\end{pmatrix}.
\end{equation}
Hence, multiplying~\eqref{tsample} through by the vector $(1,0)$ yields
\begin{equation}
C_0=\sigma_1Z_1/\sqrt{W/\nu}+\hat{C}_0,
\end{equation}
which implies that $(C_0-\hat{C}_0)/\sigma_1$ has a univariate $t$-distribution with $\nu$ degrees of freedom.
It follows that
\begin{equation}
p\left(t_{0.025,\nu}<\frac{C_0-\hat{C}_0}{\sigma_1}<t_{0.975,\nu}\right)=0.95,
\end{equation}
where $t_{0.025,\nu}$ and $t_{0.975,\nu}$ are the 0.025th and 0.975th quantiles of a univariate $t$-distribution with $\nu$ degrees of freedom, respectively.
Equivalently,
\begin{equation}
p(\hat{C}_0+t_{0.025,\nu}\sigma_1<C_0<\hat{C}_0+t_{0.975,\nu}\sigma_1)=0.95,
\end{equation}
which proves that the credible interval for $C_0$ is given by~\eqref{credint_C0}.
The credible interval for $S$ given by~\eqref{credint_S} is derived similarly by multiplying~\eqref{tsample} through by the vector $(0,1)$ instead of $(1,0)$.

\section{Posterior Predictive Distribution}
\label{app:ppd}
\setcounter{equation}{0}

The posterior predictive distribution for a new shock wave velocity measurement, $p(\tilde{y}|Y)$, given in~\eqref{ppd_density}, can be derived as follows.
The future measurement, $\tilde{Y}$, is modeled as
\begin{equation}
\tilde{Y}=\xstar'\beta+\tilde{\epsilon},
\end{equation}
where $\tilde{\epsilon}\sim N(0,\sigma^2)$.
Since
\begin{equation}
\beta|\sigma^2,Y\sim N(\hat{\beta},\sigma^2(X'X)^{-1})
\end{equation}
from~\eqref{posterior_beta_given_sigmasq_Y}, it follows from multiplying through by $\xstar'$ and adding $\tilde{\epsilon}$ that
\begin{equation}
\label{tildeY_given_sigmasq_Y}
\tilde{Y}|\sigma^2,Y\sim N(\xstar'\hat{\beta},\sigma^2(1+\xstar'(X'X)^{-1}\xstar)).
\end{equation}
The posterior predictive distribution is obtained by integrating $\sigma^2$ out of the distribution in~\eqref{tildeY_given_sigmasq_Y}.
That is,
\begin{align}
p(\tilde{y}|Y)&=\int p(\tilde{y}|\sigma^2,Y)p(\sigma^2|Y)\,d\sigma^2\\
&\propto\int \frac{1}{(\sigma^2)^{1/2}}\exp\left(-\frac{(\tilde{y}-\xstar'\hat{\beta})^2}{2\sigma^2(1+\xstar'(X'X)^{-1}\xstar)}\right)\frac{1}{(\sigma^2)^{n/2}}\exp\left(-\frac{\nu s^2}{2\sigma^2}\right)\,d\sigma^2\\
&\propto \int \frac{1}{(\sigma^2)^{(n+1)/2}}\exp\left(-\frac{1}{\sigma^2}\left[\frac{(\tilde{y}-\xstar'\hat{\beta})^2}{2(1+\xstar'(X'X)^{-1}\xstar)}+\frac{\nu s^2}{2}\right]\right)\,d\sigma^2\\
&\propto \left[1+\frac{(\tilde{y}-\xstar'\hat{\beta})^2}{\nu s^2(1+\xstar'(X'X)^{-1}\xstar)}\right]^{-(n-1)/2},
\end{align}
where $p(\sigma^2|Y)$ is given in~\eqref{pdf_sigmasq_given_Y}.
The last integral is computed using the observation that the integrand is the non-normalized density of an inverse gamma distributed random variable with shape parameter $a_*=(n-1)/2$ and scale parameter, say $b_*$, given by the expression in the square brackets.
Since the integral of a probability density function over its support equals one, the last integral is proportional to $b_*^{-a_*}$.
The last expression is the non-normalized density of a univariate $t$-distribution with $\nu$ degrees of freedom, mean $\xstar'\hat{\beta}$, and scale parameter $s^2(1+\xstar'(X'X)^{-1}\xstar)$, which implies that the posterior predictive distribution is as given in~\eqref{post_pred_dist}.

\section{Bootstrap Confidence and Prediction Intervals for Shock Wave Velocity}
\label{app:bootstrap_intervals}
\setcounter{equation}{0}

\begin{figure}[h!]
  \centering
  \begin{subfigure}[b]{0.32\textwidth}
    \centering
    \includegraphics[width=\textwidth]{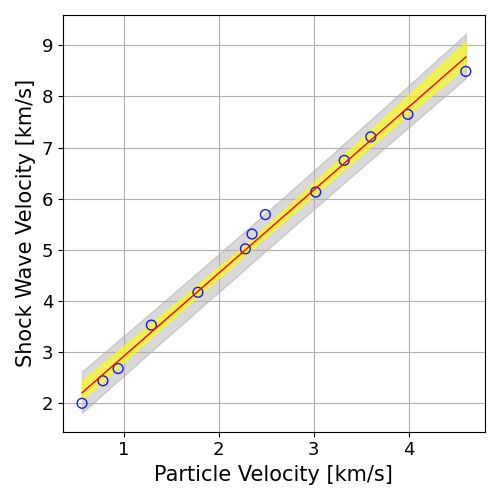}
    \caption{Argon}
  \end{subfigure}
  \begin{subfigure}[b]{0.32\textwidth}
      \centering
      \includegraphics[width=\textwidth]{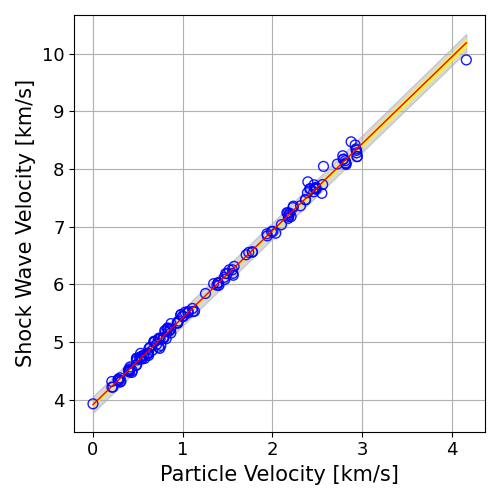}
      \caption{Copper}
  \end{subfigure}
  \begin{subfigure}[b]{0.32\textwidth}
    \centering
    \includegraphics[width=\textwidth]{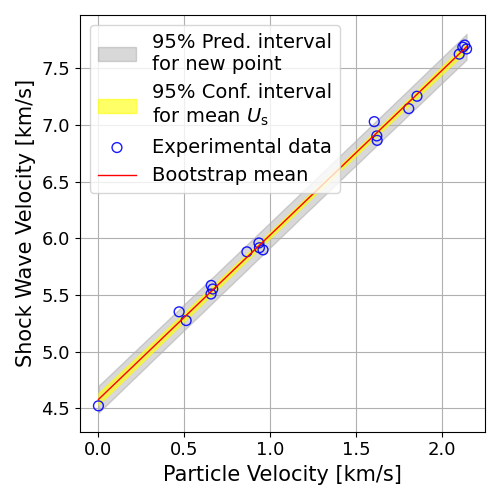}
    \caption{Nickel}
  \end{subfigure}
  \caption{Bootstrap confidence intervals for the mean shock wave velocity and prediction intervals for new shock wave velocity measurements.}
  \label{fig:bootstrap_ints}
\end{figure}

Figure~\ref{fig:bootstrap_ints} shows bootstrap confidence intervals for the mean shock wave velocity and prediction intervals for new shock wave velocity measurements.
The plot is created by first resampling the original data 100,000 times to create a large collection of bootstrap datasets.
For each bootstrap dataset, we compute $\hat{\beta}$ and then evaluate the corresponding shock wave velocities across a grid of particle velocities.
This results in 100,000 shock wave-particle velocity curves, whose mean is called the bootstrap mean in the figure.
The 95\% confidence intervals are computed as the 2.5th and 97.5th percentiles of this ensemble of shock wave-particle velocity curves.
The prediction intervals are obtained by adding normally distributed random variables with variance $s^2$ to the simulated shock wave velocity data, and then taking the 2.5th and 97.5th percentiles of the curves.
Overall, the means and intervals in Fig.~\ref{fig:bootstrap_ints} appear close to those in Fig.~\ref{fig:pred_ints}, which shows credible and prediction intervals for the posterior distribution.

\section{Informative Prior Distribution}
\label{app:informative_prior}
\setcounter{equation}{0}

In Sec.~\ref{sec:informative_prior}, examples were given of the marginal posterior distribution of $\beta$ under an informative prior distribution.
This appendix begins by deriving the joint posterior distribution of $\beta$ and $\sigma^2$, and then obtains the marginal posterior distributions of $\beta$ and of $\sigma^2$ as a by-product.

As discussed in Sec.~\ref{sec:informative_prior}, we assume a normal-inverse-gamma distribution as the prior distribution, which is written as
\begin{equation}
p(\beta,\sigma^2)=p(\beta|\sigma^2)p(\sigma^2),
\end{equation}
where $\beta|\sigma^2$ is a normal distribution with mean $\beta_0$ and covariance matrix $\sigma^2\Sigma_0$, and where $\sigma^2$ has an inverse gamma distribution with shape parameter $a_0$ and scale parameter $b_0$.
The prior mean for $\beta$ is then
\begin{align}
\Expval{\beta}&=\Expval{\Expval{\beta|\sigma^2}}\\
&=\Expval{\beta_0}\\
&=\beta_0,
\end{align}
and the prior covariance matrix is
\begin{align}
\Var{\beta}&=\Var{\Expval{\beta|\sigma^2}}+\Expval{\Var{\beta|\sigma^2}}\\
&=\Var{\beta_0}+\Expval{\sigma^2\Sigma_0}\\
&=\frac{b_0}{a_0-1}\Sigma_0,
\end{align}
since the elements of the covariance matrix of $\beta_0$ are all zero.
Hence, the prior mean and covariance information about $\beta$ can be summarized by choosing $\beta_0$, $\Sigma_0$, $a_0$, and $b_0$.
For example, decreasing the diagonal elements of $\Sigma_0$ shrinks the prior uncertainty in $\beta$.
This prior distribution is said to be informative since it biases the posterior distribution toward particular values of $\beta$ and $\sigma^2$.
The prior mean and covariance matrix could also be obtained by noting that the marginal prior distribution for $\beta$ is a $t$-distribution,
\begin{equation}
\label{prior_scale_matrix}
\beta\sim t_2\left(\beta_0,\frac{b_0}{a_0}\Sigma_0,2a_0\right),
\end{equation}
though we omit this derivation since it is similar to others given in this tutorial where $\sigma^2$ is integrated out of the joint prior distribution of $\beta$ and $\sigma^2$.

Under this prior distribution, the joint posterior distribution of $\beta$ and $\sigma^2$ is the product of posterior distributions for $\beta$ and $\sigma^2$,
\begin{equation}
p(\beta,\sigma^2|Y)\propto p(\beta|\sigma^2,Y)p(\sigma^2|Y).
\end{equation}
The posterior distribution of $\beta$, conditional on $\sigma^2$, is a normal distribution,
\begin{equation}
\label{posterior_beta_informative_prior}
\beta|\sigma^2,Y\sim N(\tilde{\beta},\sigma^2 G^{-1}),
\end{equation}
where
\begin{align}
G&=X'X+\Sigma_0^{-1}\\
\gamma&=X'Y+\Sigma^{-1}_0\beta_0\\
&=X'X\hat{\beta}+\Sigma^{-1}_0\beta_0,
\end{align}
and
\begin{equation}
\tilde{\beta}=G^{-1}\gamma.
\end{equation}
Note that the posterior mean, $\tilde{\beta}$, is a weighted average of the least squares estimate, $\hat{\beta}$, and the prior mean $\beta_0$.
The marginal posterior distribution of $\sigma^2$ is an inverse gamma distribution,
\begin{equation}
\label{posterior_sigmasq_informative_prior}
\sigma^2|Y\sim IG(\tilde{a},\tilde{b}),
\end{equation}
where the shape parameter is
\begin{equation}
\tilde{a}=a_0+\frac{n}{2}
\end{equation}
and the scale parameter is
\begin{equation}
\tilde{b}=b_0+\frac{1}{2}[Y'Y+\beta_0'\Sigma^{-1}_0\beta_0-\gamma'G^{-1}\gamma].
\end{equation}
The derivation of this result is standard,
\begin{align}
p(\beta,\sigma^2|Y)&\propto p(Y|\beta,\sigma^2)p(\beta|\sigma^2)p(\sigma^2)\\
&\propto \frac{1}{(\sigma^2)^{n/2}}\exp\left(-\frac{1}{2\sigma^2}(Y-X\beta)'(Y-X\beta)\right)\frac{1}{\sigma^2}\exp\left(-\frac{1}{2\sigma^2}(\beta-\beta_0)'\Sigma^{-1}_0(\beta-\beta_0)\right)\\
&\phantom{{}\propto{}} \frac{1}{(\sigma^2)^{a_0+1}}\exp\left(-\frac{b_0}{\sigma^2}\right)\\
&\propto \frac{1}{\sigma^2}\exp\left(-\frac{1}{2\sigma^2}\left[\beta'X'X\beta-2\beta'X'Y+\beta'\Sigma^{-1}_0\beta-2\beta'\Sigma^{-1}_0\beta_0\right]\right)\\
&\phantom{{}\propto{}} \frac{1}{(\sigma^2)^{\tilde{a}+1}}\exp\left(-\frac{1}{\sigma^2}[b_0+\frac{1}{2}(Y'Y+\beta_0'\Sigma^{-1}_0\beta_0)]\right)\\
&\propto \frac{1}{\sigma^2}\exp\left(-\frac{1}{2\sigma^2}[\beta'G\beta-2\beta'\gamma]\right)\frac{1}{(\sigma^2)^{\tilde{a}+1}}\exp\left(-\frac{1}{\sigma^2}[b_0+\frac{1}{2}(Y'Y+\beta_0'\Sigma^{-1}_0\beta_0)]\right)\\
&\propto \left[\frac{1}{\sigma^2}\exp\left(-\frac{1}{2\sigma^2}(\beta-\tilde{\beta})'G(\beta-\tilde{\beta})\right)\right]\left[\frac{1}{(\sigma^2)^{\tilde{a}+1}}\exp\left(-\frac{\tilde{b}}{\sigma^2}\right)\right].
\end{align}
Note that this posterior distribution is the product of a normal distribution for $\beta$ and an inverse gamma distribution for $\sigma^2$, and is an example of the normal-inverse-gamma distribution mentioned earlier.
Furthermore, as noted in Sec.~\ref{sec:informative_prior}, the posterior distribution under the normal-inverse-gamma prior converges to the posterior distribution under a non-informative prior as $a_0\rightarrow-1$, $b_0\rightarrow0$, and $\Sigma_0^{-1}$ converges to the zero matrix.

The marginal posterior distribution of $\beta$ under the informative prior is derived as
\begin{align}
p(\beta|Y)&\propto\int p(\beta,\sigma^2|Y)\,d\sigma^2\\
&\propto \int \frac{1}{(\sigma^2)^{\tilde{a}+2}}\exp\left(-\frac{1}{\sigma^2}\left[\tilde{b}+\frac{(\beta-\tilde{\beta})'G(\beta-\tilde{\beta})}{2}\right]\right)\;d\sigma^2\\
&\propto \left[\tilde{b}+\frac{(\beta-\tilde{\beta})'G(\beta-\tilde{\beta})}{2}\right]^{-(2\tilde{a}+2)/2}\\
&\propto \left[1+\frac{(\beta-\tilde{\beta})'G(\beta-\tilde{\beta})}{2\tilde{b}}\right]^{-(2\tilde{a}+2)/2}\\
&\propto \left[1+\frac{(\beta-\tilde{\beta})'[\frac{\tilde{b}}{\tilde{a}}G^{-1}]^{-1}(\beta-\tilde{\beta})}{2\tilde{a}}\right]^{-(2\tilde{a}+2)/2},
\end{align}
where the integral is obtained using the fact that the probability density function of the inverse gamma distribution integrates to one.
This implies that the marginal distribution of $\beta$ is a bivariate $t$-distribution with mean $\tilde{\beta}$, scale matrix $(\tilde{b}/\tilde{a})G^{-1}$, and $2\tilde{a}$ degrees of freedom, or
\begin{equation}
\label{marginal_posterior_beta_informative_prior}
\beta|Y\sim t_2\left(\tilde{\beta},\frac{\tilde{b}}{\tilde{a}}G^{-1},2\tilde{a}\right).
\end{equation}
The marginal posterior distribution of $\beta$ under the non-informative prior derived in~\ref{app:marginal_posterior_beta} is obtained as $a_0\rightarrow-1$, $b_0\rightarrow0$, and as the elements of $\Sigma_0^{-1}$ converge to zero.
Samples from this distribution can be obtained as described in Sec.~\ref{sec:uncprop}.

The parameters used to generate Fig.~\ref{fig:posterior_beta_informative} are given in Table~\ref{tab:informative_prior_hyperparameters}, where
\begin{equation}
\Sigma_0=
\begin{pmatrix}
\sigma_{0,C_0}^2 & \rho_0\,\sigma_{0,C_0}\sigma_{0,S} \\
\rho_0\,\sigma_{0,C_0}\sigma_{0,S} & \sigma_{0,S}^2
\end{pmatrix}.
\end{equation}
The prior scale matrix can then be calculated from~\eqref{prior_scale_matrix}.

\begin{table}[ht]
\caption{Parameters of the informative, normal-inverse-gamma prior distributions used to make Fig.~\ref{fig:posterior_beta_informative}.}
\label{tab:informative_prior_hyperparameters}
\centering
\begin{tabular}{|l|c|c|c|c|c|c|}
\hline
\textbf{Material} & $\beta_0'$ & $\sigma_{0,C_0}$ & $\sigma_{0,S}$ & $\rho_0$ & $a_0$ & $b_0$ \\
\hline
Argon  & $(1.32, 1.50)$ & 0.2 & 0.3 & $0$    & 5 & 0.5 \\
\hline
Copper & $(3.80, 1.62)$ & 0.2 & 0.2 & $-0.2$ & 5 & 0.5 \\
\hline
Nickel & $(4.70, 1.55)$ & 0.8 & 0.8 & $-0.8$ & 5 & 0.5 \\
\hline
\end{tabular}
\end{table}

\newpage

\bibliographystyle{apalike}
\bibliography{ms}

\end{document}